TITLE

Injection Locking and Noise Reduction of Resonant Tunneling Diode Terahertz Oscillator

AUTHORS

Tomoki Hiraoka[1, a)], Takashi Arikawa[1], Hiroaki Yasuda[2], Yuta Inose[1], Norihiko Sekine[2], Iwao Hosako[2], Hiroshi Ito[3] and Koichiro Tanaka[1, a)]

AFFILIATIONS

[1] Department of Physics, Graduate School of Science, Kyoto University, Sakyo-ku, Kyoto 606-8502, Japan

[2] National Institute of Information and Communications Technology, 4–2–1 Nukui-kitamachi, Koganei, Tokyo 184–8795, Japan

[3] Center for Natural Sciences, Kitasato University, Minami-ku, Sagamihara 252-0373, Japan

a) Author to whom correspondence should be addressed:
hiraoka.tomoki.68w@st.kyoto-u.ac.jp and kochan@scphys.kyoto-u.ac.jp

**ABSTRACT**

We studied the injection-locking properties of a resonant-tunneling-diode terahertz oscillator in the small-signal injection regime with a frequency-stabilized continuous THz wave. The linewidth of the emission spectrum dramatically decreased to less than 120 mHz (HWHM) from 4.4 MHz in the free running state as a result of the injection locking. We experimentally determined the amplitude of injection voltage at the antenna caused by the injected THz wave. The locking range was proportional to the injection amplitude and consistent with Adler's model. As increasing the injection amplitude, we observed decrease of the noise component in the power spectrum, which manifests the free-running state, and alternative increase of the injection-locked component. The noise component and the injection-locked component had the same power at the threshold injection amplitude as small as $5\times10^{-4}$ of the oscillation amplitude. This threshold behavior can be qualitatively explained by Maffezzoni's model of noise reduction in general limit-cycle oscillators.

# I. INTRODUCTION

Compact and stable terahertz (THz) sources are highly desired for future THz imaging and THz wireless telecommunication systems. THz oscillators with resonant tunneling diodes (RTDs) are good candidates for compact THz sources that can operate at room temperature.[1-4] Many advances have led to the fundamental oscillation frequency now being up to 1.98 THz.[5] The output power has also been increased; it is up to 0.4 mW at 530 - 590 GHz for a single oscillator[6] and 0.73 mW at 1 THz for a large-scale array.[7]

One of the major concerns in regard to the RTD THz oscillator is its large linewidth in the free-running state. It is typically 10 MHz for an oscillator operating around several hundred GHz,[8,9] where the statistical property and the origin of the noise have yet to be determined. Applications such as communications and RADAR require a narrow linewidth and frequency tunability. They also require the oscillator to synchronize with a frequency reference in order to perform homodyne or heterodyne detection.

The most commonly used methods to stabilize the frequency of the oscillators are injection locking[10-14] and phase-locked loop (PLL),[15] which have complementary properties.[16] PLL has an advantage in controlling the long-time frequency drift, but it is difficult to suppress the high-frequency noise faster than the loop-propagation delay time. Injection locking can often achieve the suppression of the high-frequency noise in a reasonable injection condition. However, if the free-running frequency drifts far away from the injection frequency, it is difficult to keep the injection locking. For the RTD THz oscillators, an intensive study on spectral narrowing by PLL is already reported.[17]

The injection locking of the RTD THz oscillator has been discussed in the context of sensitive THz-wave detection.[18,19] While these studies including subharmonic injection locking[20] revealed some properties such as the locking range in the middle- or large-signal regime, thorough investigation of the injection locking itself especially in the small-signal regime is missing. The small-signal regime, where the injection signal amplitude is much smaller than the oscillation amplitude, is practically important because injection-locking high-power slave oscillators by a stable, low-power master oscillator is required in most cases. Experiments in the small-signal regime is also essential to gain intuitive understanding of the locking mechanism based on simple analytical models.[10-14] This cannot be done in the middle- or large-signal regime since a simple modelling is not available and we have to rely on a numerical simulation to describe complex locking behavior.[10]

However, it has been difficult to observe intrinsic injection-locking properties of RTD THz oscillators in the small-signal injection regime. This is because an inevitable small return light always exists and affects the RTD THz oscillator, resulting in a complex behavior. This may have limited the previous studies on the injection locking in the middle-or large-signal injection regime to have a well-defined locking behavior.

In this study, we investigated the injection-locking properties of the RTD THz oscillator in the small-signal injection regime without an optical feedback effect. We used isolators for THz waves[21] to reduce optical feedback from the detection system. In the visible frequency range, such devices are well established and used in injection-locking experiments.[22-24] In the THz frequency range, such devices are still under development.[21] To observe the noise reduction behavior by injection locking, we stabilized continuous THz sources to have a linewidth below 120 mHz and utilized them for precise spectroscopy and injection locking. We performed measurements and analysis to obtain the spectra in which the free-running frequency of the RTD THz oscillator exactly equals to the injection frequency. This approach enabled us to obtain noise spectra which can be analyzed with a simple theory.[25]

We found that the injection locking caused the linewidth of the emission spectrum to decrease dramatically. We determined the amplitude of the injection voltage at the antenna of the oscillator caused by the injection THz wave. The locking range was proportional to the injection amplitude and consistent with Adler's model. As increasing the injection amplitude, the injection-locked component in the power spectrum gradually increased whereas the noise component, which manifests the free-running state, alternatively decreased. The noise reduction and injection-locking behavior can be qualitatively explained by Maffezzoni's model for general limit-cycle oscillators.[25]

## II. HETERODYNE THZ DETECTION AND INJECTION LOCKING SYSTEMS

We have constructed a heterodyne spectral-measurement system and an injection-locking system operating in the THz frequency region. Figure 1 shows the schematic setup for (a) the heterodyne detection system and (b) the injection-locking system.

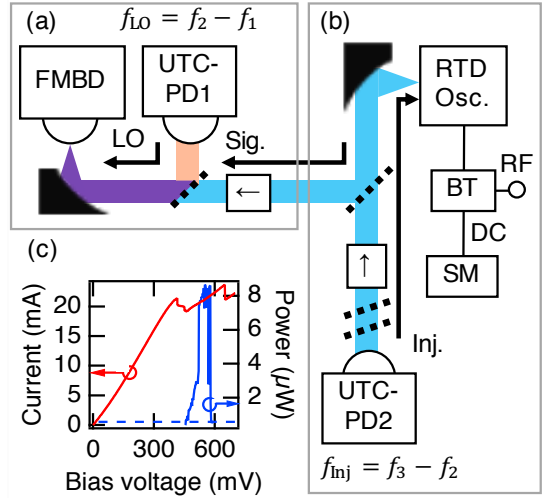

**Fig. 1** System for measuring the emission power spectrum of injection-locked RTD THz oscillator composed of (a) heterodyne detection system and (b) injection-locking system. The dotted lines represent WGs. The boxes with arrows "←" represent the isolators. SM is the source meter, and BT is a bias tee, respectively. (c) Basic properties of the RTD THz oscillator: current-voltage curve (red line) and voltage dependence of the emission power (blue line). The dashed line is the noise floor of the THz power measurement.

## A. Generation of narrow-band THz waves

We generated continuous THz waves for the local oscillator (LO) signal and the injection signal with photomixers called uni-traveling-carrier photodiodes[26] (UTC-PDs, IOD-PMAN-13001, NTT electronics, Co). We prepared three semiconductor lasers whose frequencies are stabilized to the independent frequency-comb lines separated by integer multiples of 100 MHz. We set their frequencies ($f_1$, $f_2$, $f_3$) so that their differences would each be the desired THz frequencies, i.e., the LO frequency $f_{\text{LO}} = f_2 − f_1$, and the injection frequency $f_{\text{inj}} = f_3 − f_2$. The linewidths of the THz waves were less than 120 mHz. In the locking range experiments (section IV), we had to sweep the injection frequency with an interval of less than 100 MHz. Only in that case did we stabilize the laser frequencies using a wavelength meter (Ångstrom WS7/30 IR, HighFinesse GmbH); this resulted in a linewidth of a few hundred kHz. We describe details of the narrow-band THz wave generation in Supplementary Section 1.

## B. Specifications of the heterodyne detection system

In the heterodyne detection system, the THz LO signal was generated by UTC-PD1 in Fig. 1. The LO signal and the emission of the RTD THz oscillator were combined using a wire-grid polarizer (WG). The typical power of the LO signal reflected by the WG was 10 μW. The mixed signal was detected by a Fermi-level managed barrier diode (FMBD),[27] which can detect a 0.2 - 1 THz signal with a 10-GHz intermediate frequency bandwidth. The spectrum was measured with a spectrum analyzer (MXA 9020B, Keysight Technologies Inc). Its frequency range was 10 Hz to 26.5 GHz. It can operate in real-time spectrum analyzer (RTSA) mode, in which we can capture the signal without dead time and obtain a spectrogram, which is a series of spectra over time. The resolution bandwidth (RBW) of the spectrum analyzer in RTSA mode was 240 mHz at best. The spectral resolution of the system with a frequency-comb-based THz source was limited by the RBW. The total noise floor of the detection system was determined by the background noise of the FMBD. Details of the characterization of the detection system are presented in Supplementary Section 1.

## C. Injection-locking system with isolators

Another continuous THz wave was generated with UTC-PD2 (see Figure 1) and used for injection locking. The injection power was changed using a pair of WGs. In the weaker region we also changed the laser intensity incident on UTC-PD2. This is because the extinction ratio of the WGs was not high enough to reduce the amplitude of injection field precisely down to $10^{-3}$. The THz wave emitted from the RTD THz oscillator was detected by the heterodyne detection system. We used two home-made isolators for THz waves, originally proposed by Shalaby[21] to eliminate optical feedback as shown in Fig. 1. The structure and specifications of the isolators are described in Supplementary Section 2. Without the isolators, it would have been impossible to examine the injection-locking properties in the small-signal injection regime, since there were unexpected return light from the FMBD and UTC-PD2. The RTD THz oscillator was seriously affected even by such small return light, as described in Supplementary Section 3. The effect of the return light competes with that of the externally injected signal and results in a complex behavior, especially in the small-signal injection regime. We also used electromagnetic-wave absorption sheets in sub-THz Bands[28] (Maxell, Ltd.) to eliminate unexpected return light from surrounding objects.

**D. RTD THz oscillator**

We purchased a prototype RTD THz oscillator from Rohm Co., Ltd. The oscillator had a half-wavelength dipole antenna with an RTD at the gap and a plastic leaded chip carrier package shown in Reference 29. A source meter supplied DC bias voltage to the RTD THz oscillator via the low-frequency (DC) port of a bias tee. The high-frequency (RF) port of the bias tee is used in the measurements of injection amplitude in Section IV. The current and emission power as function of bias voltage are plotted in Fig. 1 (c). Typical emission power was about 8 μW. Details of the emission power measurement are described in Supplementary Section 4.

## III. SPECTRAL FEATURES OF RTD THZ OSCILLATOR

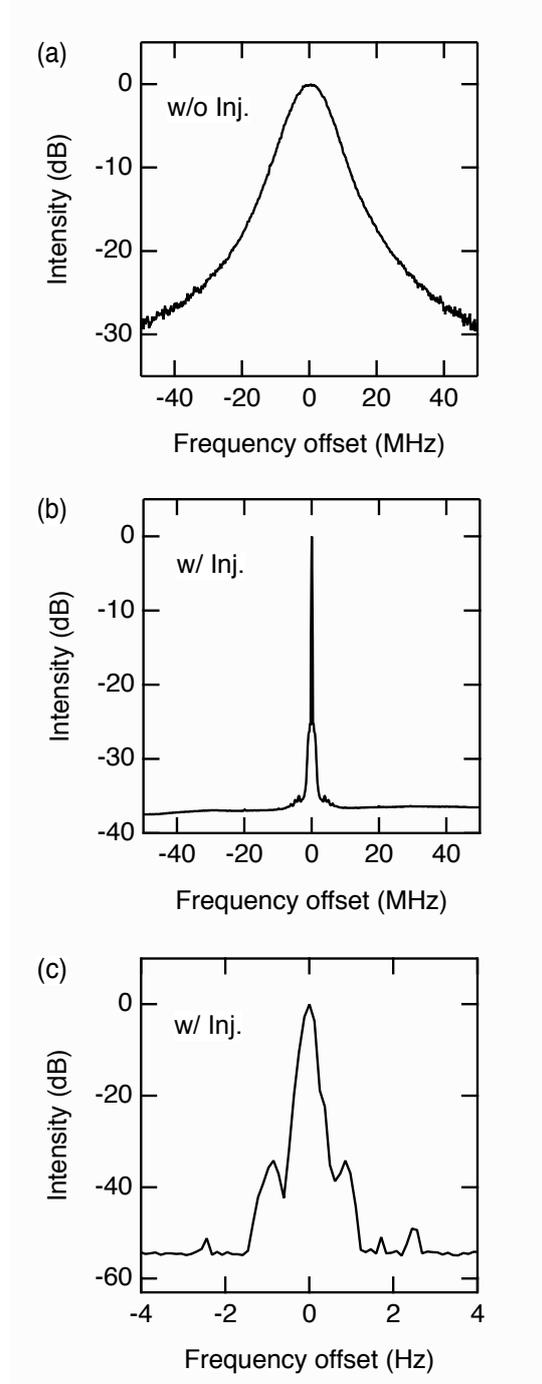

**Fig. 2** Typical emission power spectra of RTD THz oscillator. The horizontal axis is the frequency offset from the free-running frequency (322 GHz), and the vertical axis is the normalized intensity. The bias voltage was 540 mV. The LO and injection signal were frequency-stabilized with a frequency comb. We used the spectrum analyzer in RTSA mode. (a) Power spectrum without the injection signal (in the free-running state) measured with an RBW of 240 kHz. To eliminate the effect of the fluctuation, we analyzed the spectrogram with the post-selection method described in Supplementary Section 5. (b) Power spectrum with the injection signal (in the injection-locked state) measured with an RBW of 240 kHz. (c) Power spectrum of the injection-locked RTD THz oscillator measured with an RBW of 240 mHz. For Figure (b) and (c), we did not use the post-selection analysis. The noise level difference in (b) and (c) is due to the RBW difference.

### A. Free-running state

A typical emission power spectrum of the RTD THz oscillator in the free-running state is shown in Fig. 2 (a). The center frequency was about 322 GHz. We found that the center frequency fluctuates slowly in time. In order to eliminate the fluctuation and to obtain instantaneous linewidth, we applied a post-selection method; we used the spectrum analyzer in RTSA mode to acquire a spectrogram and selected only the spectra with the same center frequency. The details of the frequency fluctuation, its impact on the spectral shape and the post-selection method are described in Supplementary Section 5. The half width at half maximum (HWHM) of the free-running spectrum is 4.4 MHz. This value is on the same order as that in the previous study.[8]

### B. Injection-locked state

The emission power spectrum of the RTD THz oscillator under signal injection is shown in Fig. 2 (b). The injected power was about 2 μW in front of the RTD THz oscillator, which is the maximum power in our experiment. In this case, the oscillator was stably locked and we did not observe any effect of the fluctuation. Hence, we did not apply the post-selection analysis. The linewidth was dramatically decreased by the signal injection. Figure 2 (c) shows the injection-locked spectrum measured with an RBW of 240 mHz. The observed linewidth was 120 mHz, which is limited by the RBW. The shape of the injection-locked spectrum is almost the same as that of the injection THz signal shown in Supplementary Section 1. The small sidebands in Figs. 2 (b) and 2 (c) come from the spectrum of the LO signal and the injection signal, which is shown in Supplementary Section 1. This is consistent with a theory in which the injection-locked spectrum would be almost the same as that of the injection signal when the injection signal is strong enough.[30]

## IV. LOCKING RANGE

We swept the injection frequency and measured the emission power spectra of the RTD THz oscillator (Fig. 3). The blue trace in the top shows the free-running spectra with a vertical dashed line indicating the position of the free-running frequency. The narrow peaks in other traces (grouped by A, B, and C) show the positions of the injection frequency. As the injection frequency approaches the free-running one (group A), we start to see equally spaced sidebands. This state is called injection-pulled state[12] and commonly observed in nonlinear oscillators. As we further decrease the injection frequency below a certain point ($f_{L,max}$, red dotted line), sidebands disappear and only a narrow peak at the injection frequency remains, i.e., the RTD THz oscillator is injection locked (group B). To distinguish the injection-locked state from the injection-pulled state, we used the same criterion as a previous study[24]: if the height of the peak at the injection frequency is more than 20 dB larger than that of the sidebands, it is an injection-locked state. When the injection frequency becomes less than $f_{L,min}$ shown as a red dotted line, sidebands appear (group C); the oscillator is no more injection locked but injection pulled again. The spacing of the sidebands increases as the injection frequency depart from the free-running frequency. The frequency range from $f_{L,min}$ to $f_{L,max}$ is the locking range.

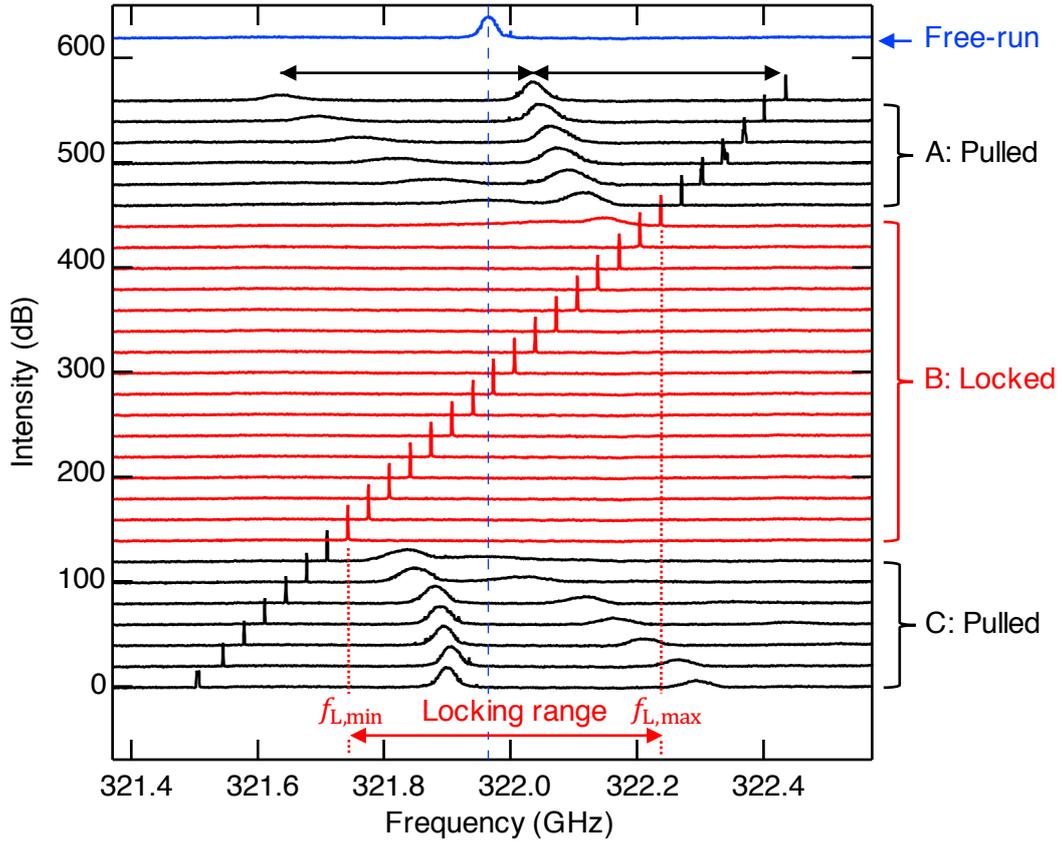

**Fig. 3** Emission power spectra of the RTD THz oscillator under various injection conditions. The free-running frequency is shown as the blue dashed line. The narrow peaks in other traces (grouped by A, B, and C) show the positions of the injection frequency. The spectra in group A and C show injection-pulled state, and the spectra in group B show injection-locked state. The bias voltage was 540 mV, and the injection power was about 2 μW, which is the maximum power in our setup. To capture the wide frequency span of 1.2 GHz, we used the spectrum analyzer in swept spectrum analyzer mode, not in RTSA mode where the frequency span is limited to 160 MHz. Although we could not apply the post-selection analysis in this experiment, the center frequency fluctuation (~ 10 MHz) does not largely affect the locking range of several hundred MHz. The RBW was 400 kHz, and each spectrum was averaged over 1 second. We stabilized the frequency of the LO and injection signal with the wavelength meter.

We measured the locking range for various amplitudes of injection field. Figure 4 (a) shows *the Arnold tongue*, the region where the injection locking occurs in the injection-frequency and injection-amplitude plane. The vertical axis is the normalized injection amplitude $k = V_{inj}/V_{inj,max}$. Here, $V_{inj}$ is the amplitude of injection voltage at the antenna caused by the injection electric field. $V_{inj,max}$ is 0.41 mV, which is the maximum value of $V_{inj}$ in the series of experiments in this paper. We determined $V_{inj,max}$ and $V_{inj}$ by making injection-amplitude measurements based on square-law detection, as described in Supplementary Section 6. According to Adler's model,[10-12] the half locking range $a_f$ of a weakly nonlinear oscillator under small-signal injection is represented as

$$a_f = \frac{f_0}{2Q}\frac{V_{inj}}{V_{osc}} = \frac{f_0 r}{2Q} \quad (1)$$

where $f_0$ is the free-running frequency, $Q$ is the Q-factor of the resonator, and $V_{osc}$ is the amplitude of oscillation voltage at the antenna. The factor $r = V_{inj}/V_{osc}$ is commonly called the injection ratio. Equation (1) predicts that the locking range is proportional to the injection amplitude, and we can see the proportionality in Fig. 4 (a). We can calculate the half locking range $a_f$ for a normalized injection amplitude $k$ of 0.75 as 400 MHz, which is on the same order as the experimentally obtained $a_f$ of 250 MHz. Parameters used in the calculation are as follows: the injection ratio $r = 1 \times 10^{-2}$ when $k = 0.75$, $f_0 = 322$ GHz, and $Q = 4$. We determined the oscillation voltage $V_{osc}$ by making emission power measurement and calculated the injection ratio $r$, as described in Supplementary Section 6. $Q$ was obtained from a finite-difference time-domain (FDTD) simulation[31] of the antenna structure on the substrate. This result would validate Adler's model for the RTD THz oscillator in the small-signal injection regime.

Figure 4 (b) shows the locking range for the normalized injection amplitude of $k = 0.75$ at several bias voltages as red vertical bars. The free-running frequency and free-running linewidth are also shown as black curves and blue dots, respectively. We can see that the locking range depends on the bias voltage. Furthermore, the locking range is asymmetric about the free-running frequency at some bias points. This behavior cannot be explained by Adler's model. It is possibly related to voltage-dependent susceptance. Hansson showed that an oscillator with a voltage-dependent susceptance may have an asymmetric locking range.[32] By conducting a circuit simulation, we confirmed that the voltage-dependent capacitance of RTD could result in a voltage-dependent and asymmetric locking range. We used a circuit model similar to the one developed by Diebold.[33] We calculated the locking range for two cases, i.e., in which the capacitance of RTD does not vary in spite of the oscillating voltage and in which it varies with the oscillating voltage. As shown in Supplementary Section 7, the latter model reproduces the voltage-dependent and asymmetric locking range. In Fig. 4 (b), the asymmetry becomes large at 555 mV, where a kink exists in the free-running frequency and a peak appears in the linewidth. We expect that the gradient of the voltage-dependent capacitance would be large here and give rise to a large asymmetry.

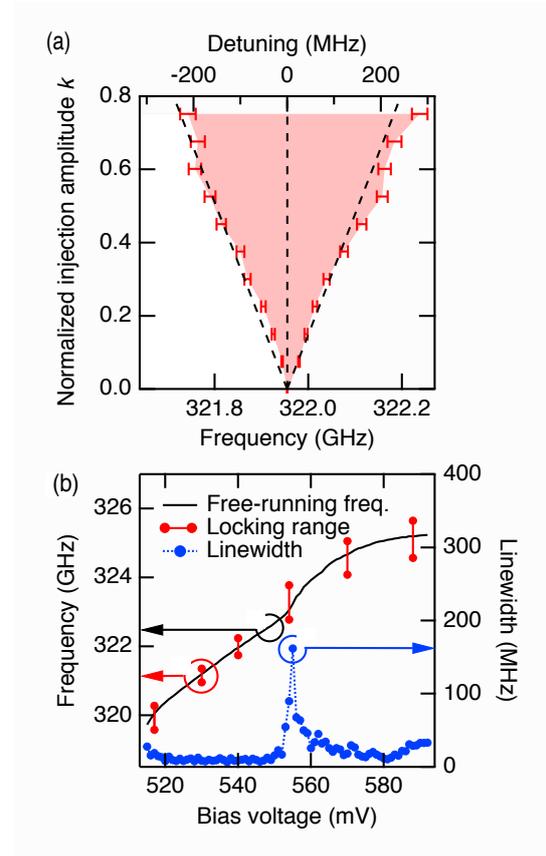

**Fig. 4** (a) Arnold tongue measured at a bias voltage of 540 mV (red filled area). The vertical axis is the injection amplitude $\boldsymbol{k}$ normalized to the maximum injection-voltage amplitude in the series of experiments in this paper. The maximum value of $\boldsymbol{k}$ in this figure is 0.75, which corresponds to the injection ratio of $\boldsymbol{r = 1 \times 10^{-2}}$. The vertical dashed line indicates the free-running frequency. The diagonal dashed lines are guides-to-the-eye indicating that the locking range is proportional to the normalized injection amplitude and symmetric about the free-running frequency. (b) Bias-voltage dependence of the free-running frequency (black curve), locking range for the normalized injection amplitude of $\boldsymbol{k = 0.75}$ (red vertical bars), and free-running linewidth (blue dots) presented in the voltage range of 515 - 592 mV. (a) and (b) was obtained from the locking-range measurements, in which we did not use the post-selection analysis.

## V. NOISE REDUCTION

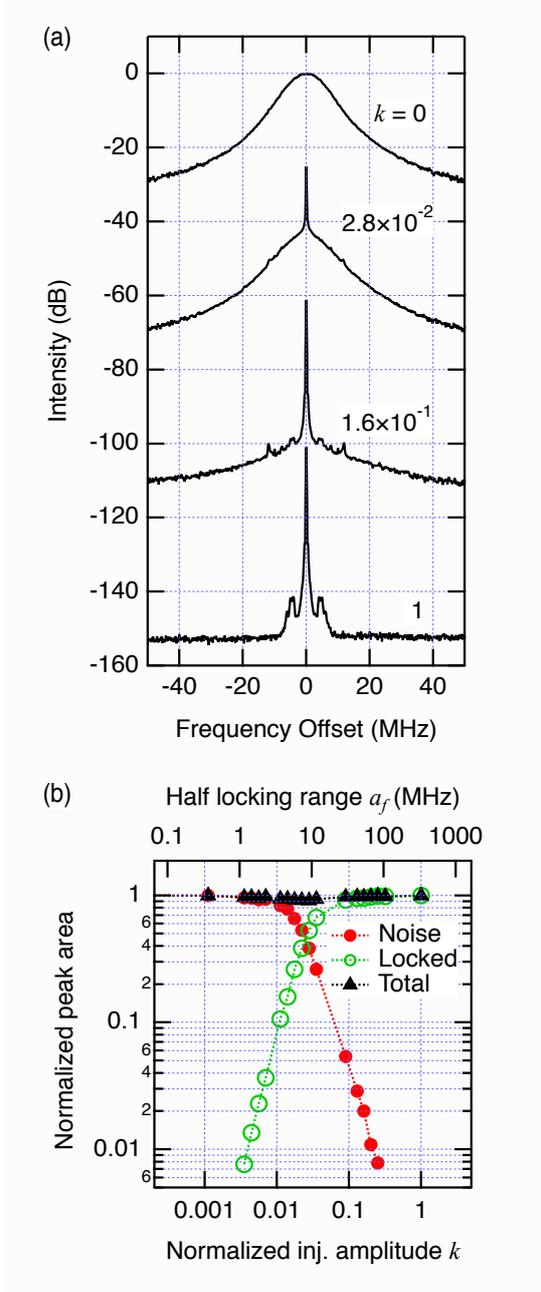

**Fig. 5** (a) Emission power spectra of the RTD THz oscillator for various normalized injection amplitudes $k$. The bias voltage was 540 mV. The frequencies of the LO and injection signal were stabilized with the frequency comb. The injection frequency was set to the free-running frequency. To eliminate the effect of the frequency fluctuation, we extracted the spectra from spectrograms by using the post-selection analysis shown in Supplementary Section 5. The RBW was 240 kHz. The sweep time for each spectrum in the spectrograms was 100 μs. (b) Dependence of the normalized peak area of the spectral components on the normalized injection amplitude $k$.

### A. Noise reduction threshold

It is important to explore the minimum injection strength to injection-lock the oscillator. We set the injection frequency at the center of the free-running spectrum and changed the amplitude of the injection field. Figure 5 (a) shows the spectral shape for various normalized injection amplitudes $k$. The free-running frequency fluctuates in time and it does not exactly coincide with the injection frequency as shown in Supplementary Section 5. This makes the analysis of the noise spectra difficult. To obtain the spectra in which the free-running frequency exactly coincides with the injection frequency, we used the post-selection method described in Supplementary Section 5. In the case of no injection ($k = 0$), there is only a broad peak with the HWHM $w_D$ of 4.4 MHz. We call the broad peak the noise component. As the normalized injection amplitude increases, the noise component starts to diminish. Instead, a narrow peak, i.e., the injection-locked component, appears and grows. In the case of $k = 1$, only the injection-locked component remains. The weak sidebands (40 dB below the main peak) at several MHz are attributed to the injection signal or the LO signal, as described in Section III B.

We derived the power of the noise component and the injection-locked component from the spectral area with the method described in Supplementary Section 8. Figure 5 (b) shows the dependence of the normalized peak area on the normalized injection amplitude $k$. The top axis shows the estimated half locking range $a_f$ for each normalized injection amplitude. For the estimation, we performed an interpolation and an extrapolation of the locking range shown in Fig. 4 (a) based on Eq. (1) of Adler's model. As the normalized injection amplitude $k$ increases, the noise component decreases and the injection-locked component increases. The total power, i.e., the sum of the two components, is almost conserved. The power of the injection-locked component exceeds that of the noise component when the normalized injection amplitude $k$ is 3×10⁻², which corresponds to an injection ratio $r$ of $5 \times 10^{-4}$. At this threshold, the half locking range $a_f$ is about 9 MHz as we can see in the top axis.

To examine the threshold, we applied Maffezzoni's model,[25] which describes the noise reduction by injection locking in general nonlinear oscillators with a white noise source. The model predicts that the intensity ratio between the noise component and the injection-locked component becomes unity when $a_f = w_D/\ln 2$. This gives $a_f = 6.3$ MHz, very close to the experimentally obtained 9 MHz. Hence, the threshold value is consistent with Maffezzoni's model.

Finally, it is noteworthy that the RTD THz oscillator can be injection-locked by such a small signal which corresponds to an injection ratio $r$ of $5 \times 10^{-4}$. This would be useful in a practical situation to stabilize it with a weak injection signal. At the same time, the small threshold also implies that the RTD THz oscillator is sensitively disturbed by an external THz signal, including tiny optical feedback.[34-36] This fact points to the need for isolators in our experiment and also in the future applications of the RTD THz oscillators.

**B. Details of the noise spectra**

Figure 6 (a) shows the power spectrum of the noise component for several injection strengths in terms of a log-log plot. Here, we used the half-locking range $a_f$ at each injection strength to label the experimental condition, where $a_f = 0$ MHz is the free-running state. As the injection strength increases, the noise decreases. In the strong-injection limit ($a_f = 330$ MHz), the spectrum is almost the same as the heterodyne spectrum of the LO signal and the injection signal. Figure 6 (b) shows the power spectrum predicted by Maffezzoni's model. One can see that the frequency range where the noise reduction occurs is qualitatively reproduced by Maffezzoni's model; the noise is substantially reduced within the locking range. It should be noticed that there is a difference in the slope of the high-frequency noise between the experimental spectra ($\Delta f^{-3}$) and theoretical spectra ($\Delta f^{-2}$). We expect the slope of $\Delta f^{-3}$ in the experiment comes from a flicker noise source,[37] while Maffezzoni's model considers a white noise source. We discuss this possibility in more detail in Supplementary Section 9.

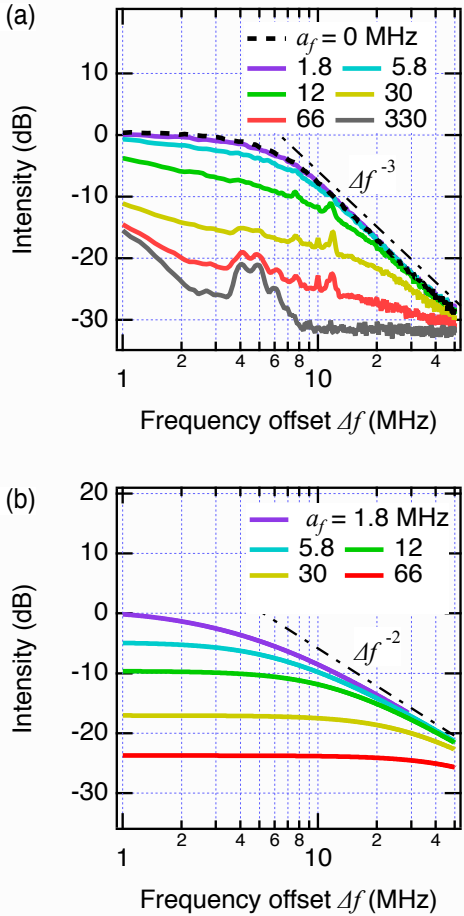

**Fig. 6** (a) Measured power spectra for various injection strengths in log-log plot. The half locking range $a_f$ represents the injection strength. The chain line is a guide-to-the-eye indicating the slope of $\Delta f^{-3}$ in the free-running spectrum. This figure is based on Fig. 5 (a), which is obtained with the post-selection analysis. (b) Calculated power spectra for various injection strengths with Maffezzoni's model. As the parameter of the theoretical curves, we used the half locking range $a_f$, the free-running half-linewidth $w_D = 4.4$ MHz, and the normalization factor in the spectral height. The chain line is a guide-to-the-eye indicating the $\Delta f^{-2}$ slope of the high-frequency noise.

## VI. CONCLUSION

We characterized the injection locking of an RTD THz oscillator to a small-amplitude continuous THz wave. We performed precise measurements with frequency-stabilized THz sources, a real-time spectrum analyzer, and THz isolators. Injection locking reduced the linewidth of the emission power spectrum from 4.4 MHz to less than 120 mHz. The dependence of the locking range on the injection ratio is consistent with Adler's model. Use of an RTSA enabled us to obtain the noise spectra which can be compared with a simple model. We determined the noise reduction threshold, setting the injection frequency at the free-running frequency. The threshold injection ratio was $5\times10^{-4}$. At the threshold, the locking range is almost the same as the free-running linewidth, as expected from Maffezzoni's model. We also applied Maffezzoni's model to the noise reduction process, which yielded a qualitative understanding of the injection-locking properties of the RTD THz oscillator.

**SUPPLEMENTARY MATERIAL**

See the supplementary material for further details.

**ACKNOWLEDGEMENTS**

The authors would like to thank M. Asada and S. Suzuki for valuable discussions. We appreciate Y. Takida and H. Minamide for their helpful discussions and for calibrating the detectors. We are grateful to Maxell holdings, Ltd. for providing us electromagnetic-wave absorption sheets in sub-THz Bands. This work was supported by JST ACCEL (Grant No. JPMJMI17F2).

**DATA AVAILABILITY**

The data that support the findings of this study are available from the corresponding author upon reasonable request.

**Supplementary information**

**Section 1   THz wave generation with UTC-PD**

Here, we describe the details of narrow-band continuous THz wave generation by differential frequency photomixing with uni-traveling-carrier photodiodes (UTC-PDs). In this method, we injected outputs from two CW lasers to a UTC-PD. The frequency difference of the lasers was set to the desired THz frequency. We stabilized the frequencies of the CW lasers using a wavelength meter (WLM) or an optical frequency comb.

Figure S1 shows the feedback configuration using a WLM (Ångstrom WS7/30 IR, HighFinesse GmbH). Merit of this method was that the center THz frequency was continuously tunable. We used three frequency-tunable laser diodes (LDs) operating at 1.5 μm, i.e., DLpro (TOPTICA Photonics AG), CTL1550 (TOPTICA Photonics AG), and ORION (RIO lasers). We used optical fibers to connect optical components such as the LDs, the optical amplifiers, the UTC-PDs, and the WLM. We could read an operating frequency and also set a target frequency for each LD using WLM. Error signals were generated in WLM and sent to DLpro and CTL to control their frequencies. ORION was operated without the feedback control. The linewidths of the LD frequencies were a few hundred kHz, which limited the linewidth of the generated THz wave.

Figure S2 (a) shows the schematic diagram for the frequency stabilization method using an optical frequency comb as a frequency standard. With this method, the linewidth of the generated THz wave was as narrow as 120 mHz, which was limited by the frequency resolution of the measurement system. The center frequency could be integer multiple of 100 MHz. For the optical frequency comb, we used a mode-locked Er-doped fiber laser (OCLS-100DP-KY, NEOARK CORPORATION). The optical frequency comb had a comb-like spectrum, and the frequencies of the lines were expressed as

$$f_n = f_{\text{CEO}} + n f_{\text{rep}} \qquad (S1-1)$$

where $n$ is an integer, $f_{\text{CEO}}$ is the carrier-envelope offset frequency, and $f_{\text{rep}}$ is the repetition frequency. In this study, $f_{\text{CEO}}$ and $f_{\text{rep}}$ were 10 MHz and 100 MHz, respectively. Both $f_{\text{CEO}}$ and $f_{\text{rep}}$ were stabilized using the 10 MHz frequency reference from atomic clocks in Global positioning satellite. To stabilize an LD, we chose one of the comb lines close to the target frequency and measured the beat frequency between the selected comb line and LD. The beat frequency is stabilized to 10 MHz with a feedback control unit (OCLS-STB-KY, NEOARK CORPORATION). We chose another comb line to stabilize another CW laser to generate THz wave with the difference frequency between the two comb lines. To generate 300 GHz signal, the difference of the indices of the comb lines ($\Delta n$) is $3 \times 10^3$. The instability of $f_{\text{rep}}$ ($\Delta f_{\text{rep}}$) was $10^{-4}$ Hz in 1 second. Hence, the expected linewidth of the 300 GHz signal is estimated to be $\Delta n \times \Delta f_{\text{rep}} \sim 300$ mHz, which is comparable to the measured value (120 mHz). Figure S2 (b) shows the feedback configuration. In this setup, we controlled all the LDs.

We characterized the spectral resolution of the heterodyne measurement system by using the LO signal and the input signal stabilized with a frequency comb. Figure S3 shows the setup for the characterization. The heterodyne detection part is the same as the one shown in Fig.1 (a). Figure S4 shows the measured power spectrum. The half-width at half maximum (HWHM) of the heterodyne spectrum is 120 mHz, which is limited by the best RBW of 240 mHz of the spectrum analyzer operated in real-time spectrum analyzer (RTSA) mode. We can see some minor sidebands 40 dB less than the carrier signal at several MHz from the center, which results from the spectra of the feedback-controlled laser. Due to the very low signal level, those sidebands do not matter in this study.

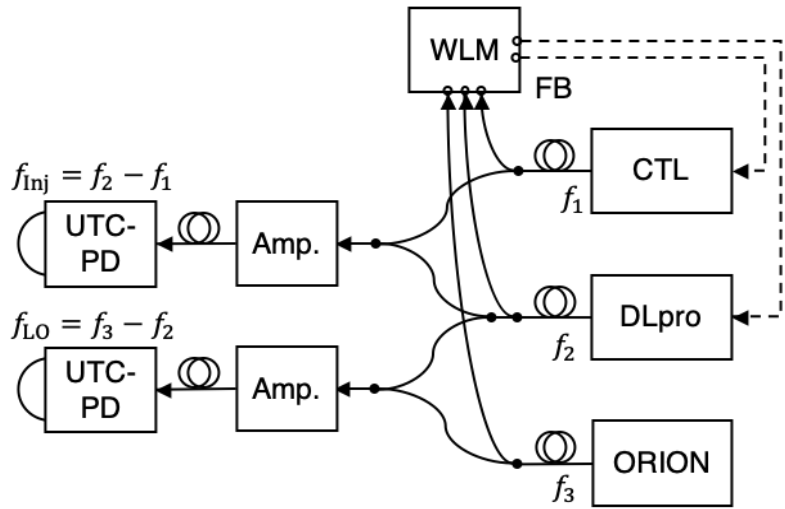

**Fig. S1** Feedback configuration with a wavelength meter. Three semiconductor lasers (CTL, DLpro and ORION) are connected to a wavelength meter (WLM) via optical fiber. Feedback control on the frequencies of CTL and DLpro was performed. Emission from CTL and DLpro are combined with a fiber coupler, amplified with an optical amplifier (Amp.), and injected to a UTC-PD to generate the injection signal. Similarly, emission from DLpro and ORION are used to generate the local oscillator.

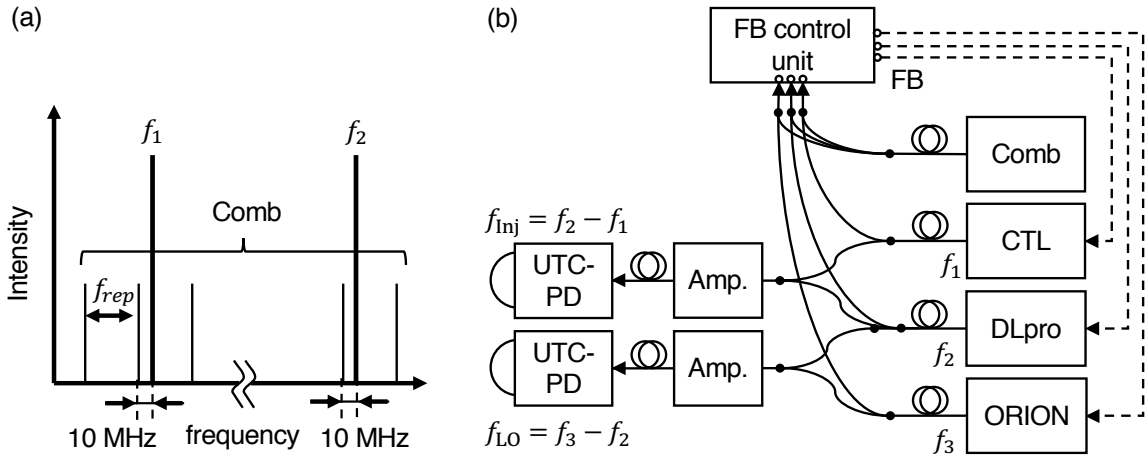

**Fig. S2** Feedback method with an optical frequency comb: (a) Schematic diagram for the stabilization method and (b) feedback configuration.

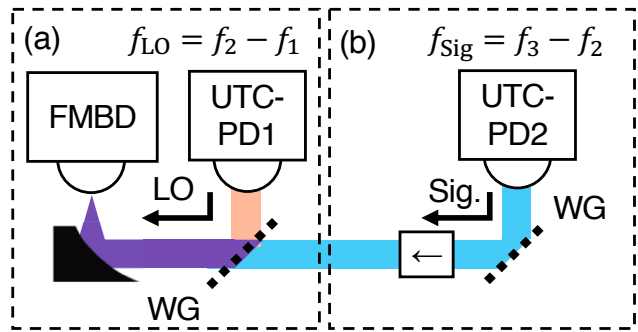

**Fig. S3** Setup for characterizing the spectral resolution consists of (a) heterodyne detection part. and (b) input signal part. We stabilized all the laser frequencies $(f_1, f_2, f_3)$ with the frequency-comb based feedback control method.

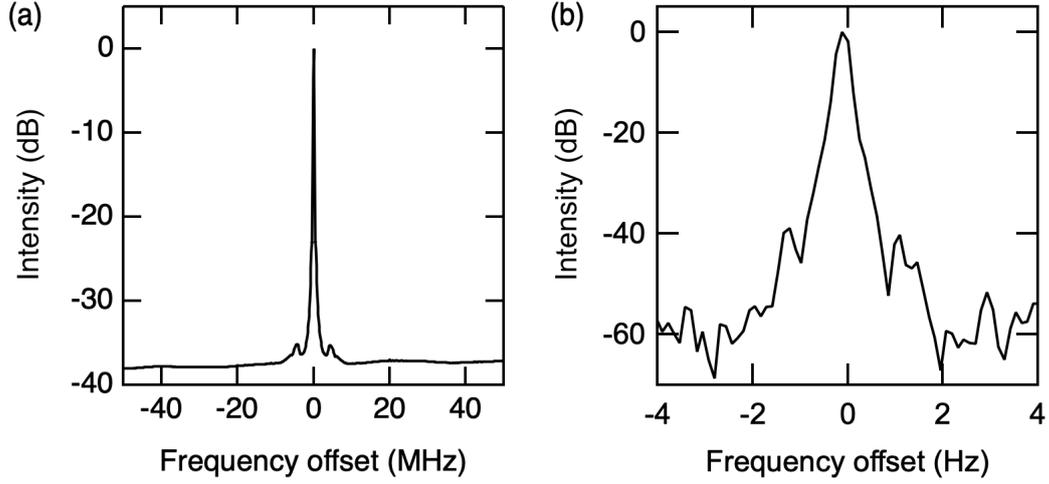

**Fig. S4** (a) Power spectrum of the frequency-comb based THz signal measured in the setup of Fig. S3. We used the spectrum analyzer in RTSA mode with (a) an RBW of 240 kHz and (b) an RBW of 240 mHz.

**Section 2  Structure and properties of THz isolators**

In this section, we describe the structure and the properties of the THz isolators. We have constructed THz isolators, which has been originally proposed by Shalaby.[1] Figure S5 is the photograph of an isolator. The isolator consists of an anisotropic Sr-ferrite magnet of approximately 2 mm-thick (Himeji Denshi Co., Ltd.) and two wire-grid polarizers (WG). The thickness of the magnet is determined to set the Faraday rotation angle to approximately 45 degrees. It was not precisely 45 degrees, and we adjusted the angle of the WGs to minimize the backward transmission. The magnet and the WGs are tilted so that the THz wave reflected at the surfaces does not return to the signal source. The isolator was constructed on a breadboard to change its position without changing the configuration inside the isolator. Specifications of two isolators at 322 GHz are summarized in Table ST1.

**Table. ST1** Transmission and polarization rotation angle of the isolators at 322 GHz.

|  | **Forward transmission** $T_f$ | **Backward transmission** $T_b$ | **Faraday rotation angle (degree)** |
|---|---|---|---|
| **Isolator 1** for injection locking | $5 \times 10^{-2}$ (-13 dB) | $7 \times 10^{-4}$ (-32 dB) | 45.0 |
| **Isolator 2** for heterodyne system | $6 \times 10^{-2}$ (-12 dB) | $6 \times 10^{-4}$ (-32 dB) | 50.5 |

The small forward transmission comes from an absorption in the Sr-ferrite magnets. Investigation of magnets without absorption loss and improvement of the forward transmission should be an important challenge in future.

Figure S6 shows schematics of the evaluation setup of the forward transmission and the backward transmission. We generated a THz wave using a UTC-PD, and measured its power using an FMBD. The transmission of isolator 2 was measured. Isolator 1 was used to avoid the formation of a standing THz wave in Figure S6 (d), which would have disturbed a precise power measurement. The WG between isolator 1 and isolator 2 was used to align the polarization of the THz beam incident on isolator 2. We modulated the input laser to the UTC-PD with an electro-optic modulator to modulate the intensity of the THz wave. Square-law detection signals from FMBD was measured with a lock-in amplifier. We calculated the forward and backward transmission of the isolator as follows:

$$T_\text{f} = \frac{P_\text{trans,f}}{P_\text{in}} \tag{S2-1}$$

$$T_\text{b} = \frac{P_\text{trans,b}}{P_\text{in}} \tag{S2-2}$$

where $P_\text{trans,f}$ is the forward transmission power, $P_\text{trans,b}$ is the backward transmission power, and $P_\text{in}$ is the incident power.

In Fig. S7, we show the details of the experimental setup of Fig. 1, especially focusing the polarization of the THz waves. The angles of the WGs are set to obtain a large signal in the heterodyne detection and high injection power. The reflection of the RTD emission from UTC-PD2 and the FMBD was eliminated by the isolators. The round-trip attenuation to the reflected field amplitude was $\sqrt{T_\text{f}T_\text{b}} = 6 \times 10^{-3}$. Several factors, such as the reflectivity at UTC-PD2 and the FMBD, would also have reduced the amplitude of the feedback field by more than one order. From these results, we expect that the amplitude ratio of the voltage at the antenna caused by the optical feedback and the oscillation voltage is less than $5 \times 10^{-4}$, which is the threshold for the injection locking. This enabled us to measure the intrinsic properties of the injection locking in the small-signal injection regime.

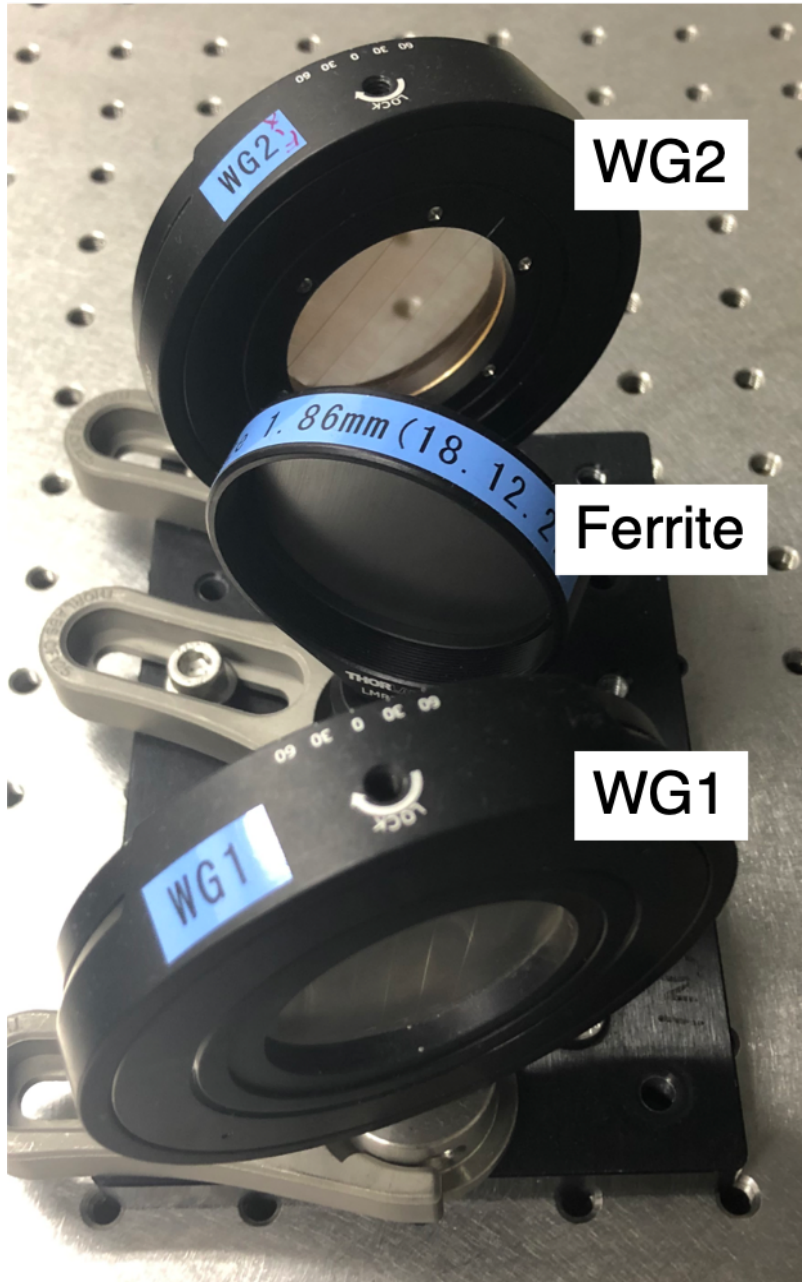

**Fig. S5** Photograph of an isolator used in the experiment. It is composed of a ferrite magnet and two wire-grid polarizers (WG1 and WG2).

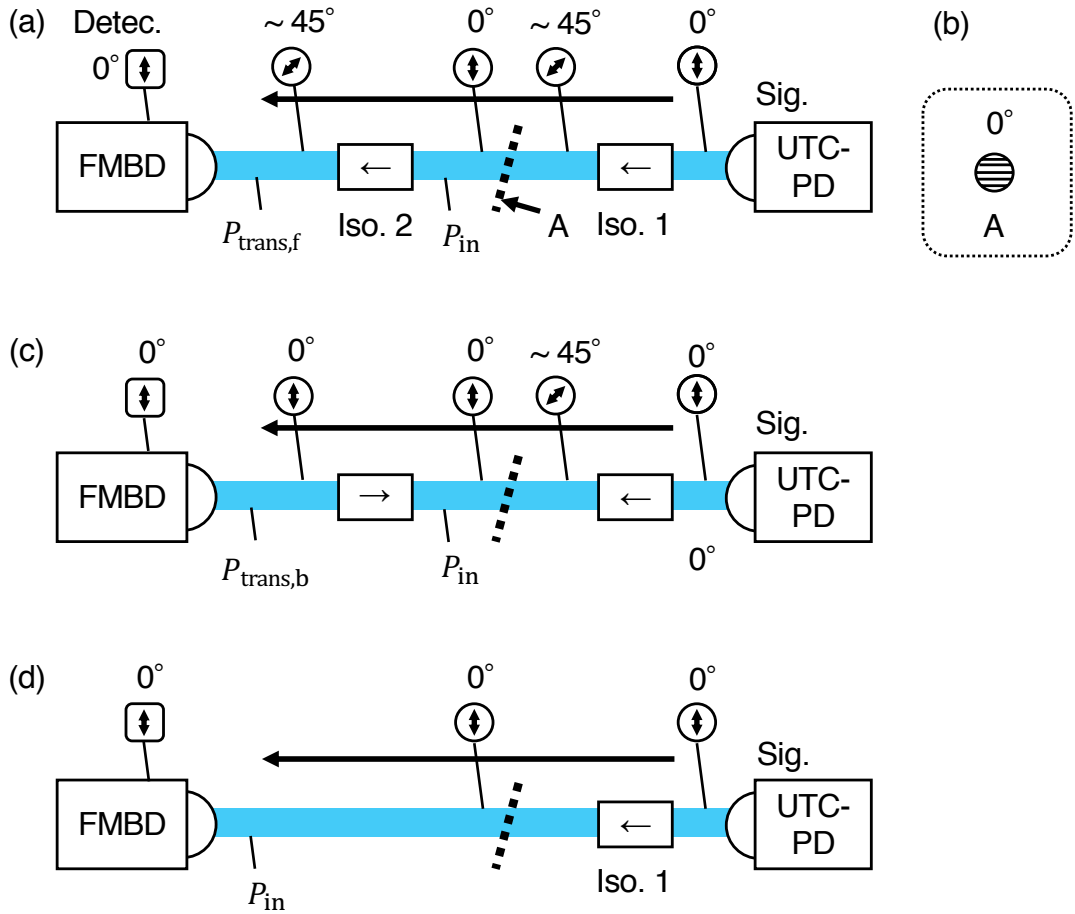

**Fig. S6** Optical system for the transmission measurement of the isolators.. (a) Setup for the forward transmission power measurement of isolator 2. $P_{in}$ is the power incident on the isolator, and $P_{trans,f}$ is the forward transmission power, respectively. The arrows in the circle indicate the polarization angle at each position, which is seen from an observer standing on the optical table and looking toward the beam propagation direction. The angle denoted is the polarization angle relative to the vertical axis. The arrow in the square represents the sensitive polarization axis of the FMBD. (b) The angle of the WG in (a) seen from the observer from point A. The direction of the wires is shown in the picture, and the angle denoted represents the transmission angle. (c) Setup for the backward transmission power ($P_{trans,b}$) measurement. (d) Setup for the incident power measurement.

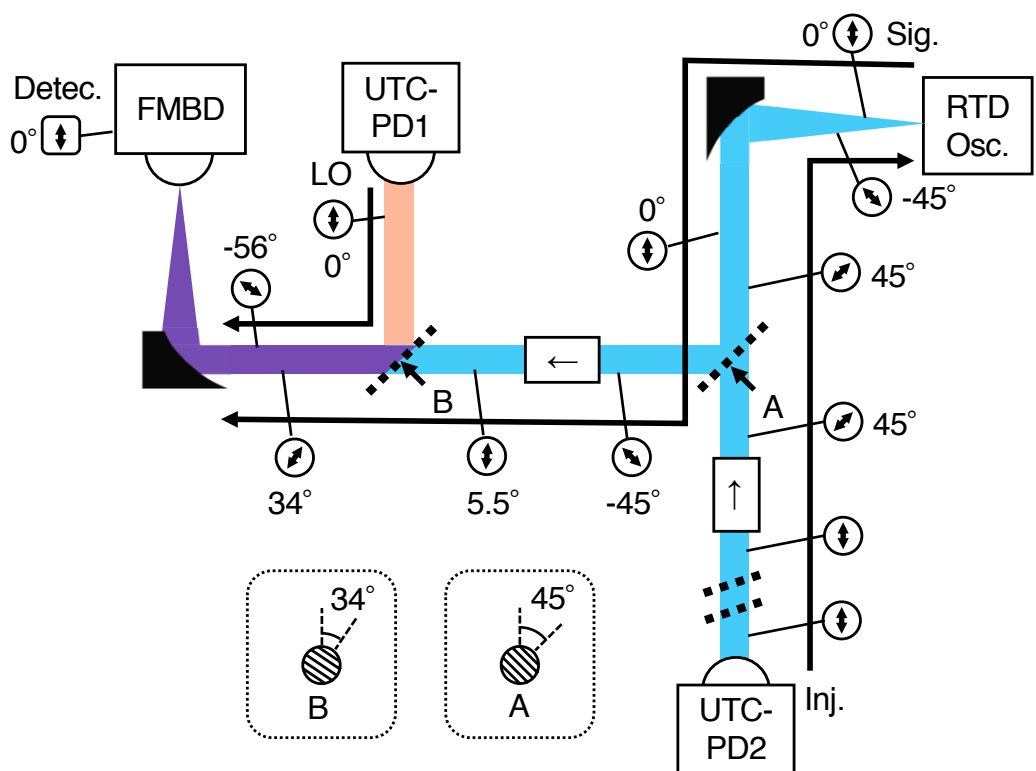

**Fig. S7** Polarization configuration in the experimental setup shown in Fig. 1. The same symbols as those of Fig. S6 represent the polarization of the THz wave, the WG angle, and the sensitive polarization axis of the FMBD.

**Section 3   Optical feedback effect on the RTD THz oscillator**

Here we describe the optical feedback effect on the RTD THz oscillator. When we remove the isolator for the detection path, a reflected THz wave of a small amplitude from the FMBD goes back to the RTD THz oscillator. We compared the emission power spectrum with and without the isolator for the detection path. We scanned the position of the RTD THz oscillator along the z-direction (relative distance: $dz$) to change the time delay of the feedback as shown in Fig. S8. Figure S9 (a) shows the emission power spectrum measured with moving the RTD THz oscillator without the isolator (a). One can see that the oscillation spectrum is largely affected by the distance $dz$. When we use the isolator (Fig. S9 (b)), the oscillation frequency is almost independent of the position of the RTD oscillator. The small change in the oscillation frequency is due to the frequency fluctuation of the RTD oscillator, as shown in the control experiment (Fig. S9 (c)). One possible explanation for the spectral change in Fig. S9 (a) is an optical feedback effect from the reflection from the FMBD, which may have formed an external cavity with a series of longitudinal modes. Systematic shift in Fig. S9 (a) should be assigned to the shift of the longitudinal modes due to the distance change between the RTD oscillator and the FMBD ($dz$).

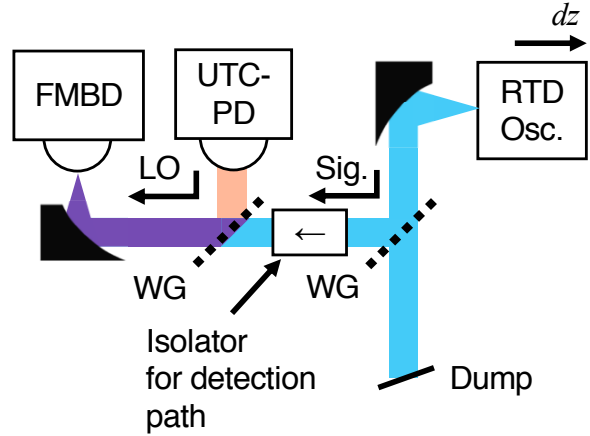

**Fig. S8** Setup for the demonstration of the optical feedback effect. The emission of the RTD THz oscillator is partially reflected by a WG and enters the heterodyne detection system. The transmitted part of the emission is dumped.

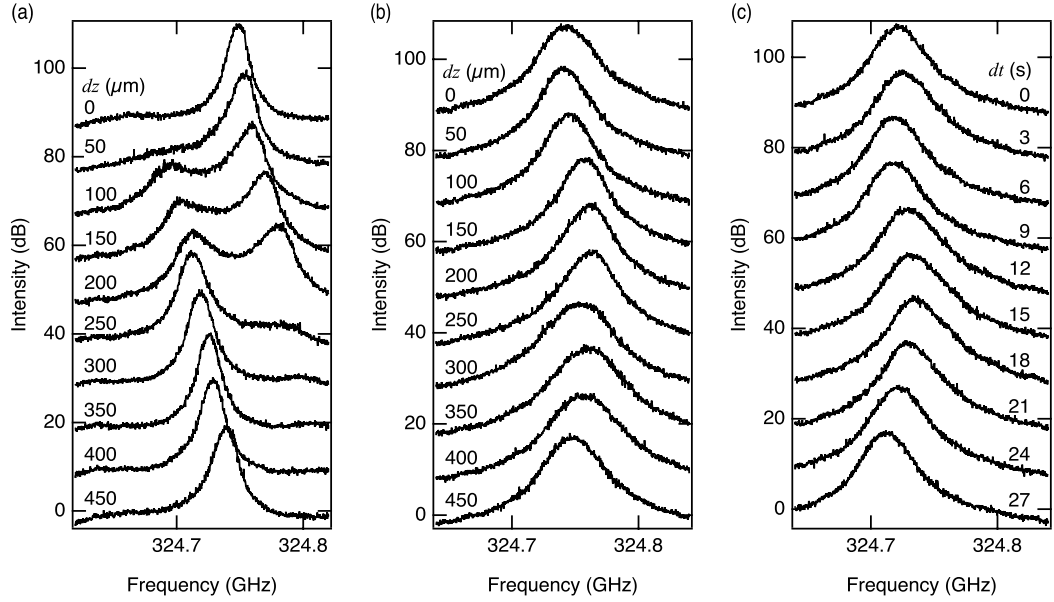

**Fig. S9** Optical feedback effect. (a) Emission power spectra of the RTD THz oscillator measured with scanning the position of the oscillator $dz$ without using the isolator. (b) Similar series of emission power spectra taken with the isolator. (c) Control experiment: Emission power spectrum of the RTD THz oscillator taken every 3 seconds without scanning the position while using the isolator. The measurement timing is denoted by $dt$. The measurement interval of 3 seconds is the same in (a), (b) and (c). To capture the wide frequency span of 200 MHz, we used the spectrum analyzer in swept spectrum analyzer mode. The RBW was 100 kHz.

**Section 4  Experimental setup for absolute output power measurement**

We measured the emission power of the RTD THz oscillator using the setup shown in Fig. S10. The emission was modulated in a square-wave on-off shape with an optical chopper. The modulation frequency was 11 Hz. The detector was a calibrated pyroelectric detector (THz 20, SLT Sensor & Lasertechnik GmbH). The detected signal was measured with a lock-in amplifier. In this measurement, we did not use an isolator because the standing THz wave that significantly affect the precision of the power measurement was not formed.

Figure 1 (c) shows the properties of the RTD THz oscillator, i.e., the voltage dependence of the current and the emission power. The current-voltage curve was measured at the feed point of the antenna. It had a negative-differential conductance region in the range of 415 mV to 650 mV. We observed a stable oscillation in the range of about 460 mV to 580 mV. We note that the lower and upper limit of the oscillation voltage range slightly changed depending on the experimental conditions. Typical emission power was 8 μW.

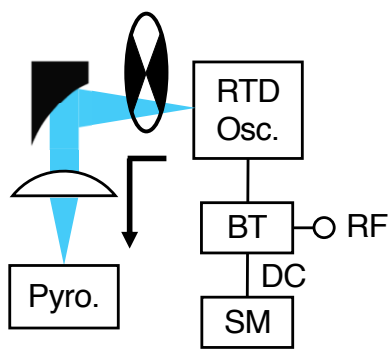

**Fig. S10** Setup to measure the emission power of the RTD THz oscillator. Pyro., BT and SM represents pyroelectric detector, bias tee and source meter, respectively.

**Section 5  Fluctuation of the free-running frequency and a post-selection method**

The center frequency of the RTD THz oscillator in the free-running state fluctuated in time. In Supplementary Section 5 (1), we show the basic properties of the fluctuation. In Supplementary Section 5 (2), we show how much the fluctuation affects the emission power spectra. In Supplementary Section 5 (3), we describe the post-selection analysis for characterize the spectra without the effect of the fluctuation. We used this method to take data shown in Fig. 2 (a), Fig. 5, and Fig. 6.

**(1) Properties of free-running-frequency fluctuation**

We measured a spectrogram, which is a series of emission power spectrum over time using the spectrum analyzer (MXA 9020B, Keysight Technologies Inc.) in RTSA mode. Figure S11 (a) shows a bare spectrogram in 1 second measured in the free-running case. The time between each trace is 115 μs. We can see fluctuation in the instantaneous center frequency $\omega_o(t)$. We derive the center frequency as the spectral centroid. Figure S11 (b) shows the temporal change of the spectral centroid of Fig. S11 (a) in terms of the offset from its average. This corresponds to the frequency noise $\delta\omega(t) = \omega_o(t) - \overline{\omega_o(t)}$, where $\overline{\omega_o(t)}$ is the center frequency averaged over time. Figure S11 (c) shows its power spectrum. We can see that the frequency noise has $1/f$ spectrum. This indicates that a parameter which affects the free-running frequency fluctuates with $1/f$ spectrum. A possible candidate for such a parameter is the capacitance of the RTD. The origin of the $1/f$ fluctuation cannot be determined from this measurement.

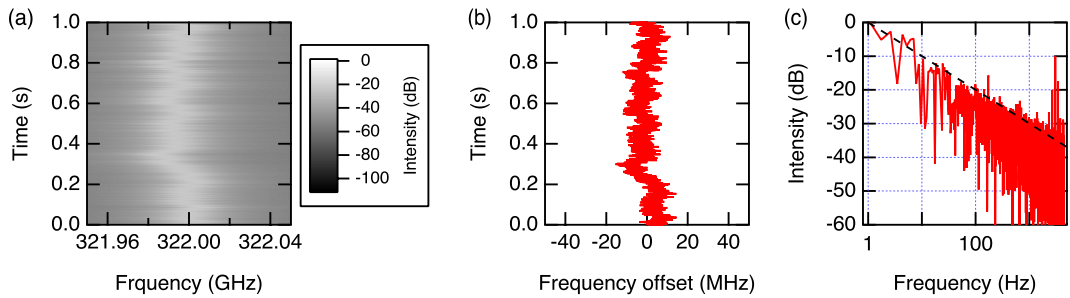

**Fig. S11** Properties of the frequency fluctuation. (a) Bare spectrogram of the emission signal measured in the free-running case. (b) Temporal change of the spectral centroid of Fig. S11 (a), i.e., frequency noise. (c) Power spectrum of the frequency noise.

**(2) Impact of the fluctuation on the spectra**

Figure S12 shows the spectrogram measured with the injection signal of several amplitudes. The horizontal axis is the frequency offset from the injection frequency. Figure S12 (a) shows the free-running case with fluctuation. Figure S12 (b) shows the spectrogram measured with very weak injection. We can see a narrow peak, i.e., injection-locked component at the injection frequency and a broad peak, i.e., noise component fluctuating in time. Figure S12 (c) shows the spectrogram taken with the maximum-amplitude injection in our setup. We can see only a narrow peak at the center with no fluctuating component. This means that the RTD THz oscillator is perfectly injection locked.

In Figure S13, we show spectra at several timings extracted from Fig. S12 to show the spectral shape clearly. As we described above, the noise component in Fig. S13 (a) and (b) fluctuates in time. We emphasize that the spectral shapes in Fig. S13 (b) are different. This is because the relation between the injection frequency and the free-running frequency is different for each trace. Hence, we need to choose spectra in which these two frequencies coincide in order to discuss the noise spectra with a simple model. If the injection signal is sufficiently strong, we observed perfectly injection-locked spectra as shown in Fig. S13 (c).

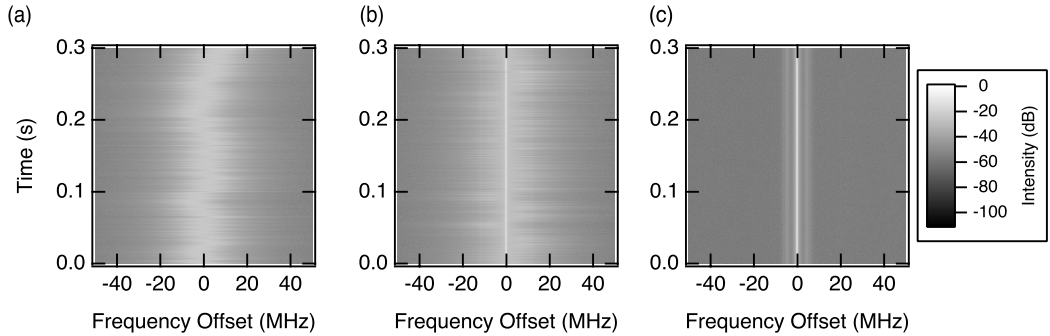

**Fig. S12** Spectrogram measured (a) with no injection signal and (b), (c) with injection signal whose frequency is close to the free-running frequency. The normalized injection amplitudes $k$ were (b) $2.8 \times 10^{-2}$ and (c) $1$.

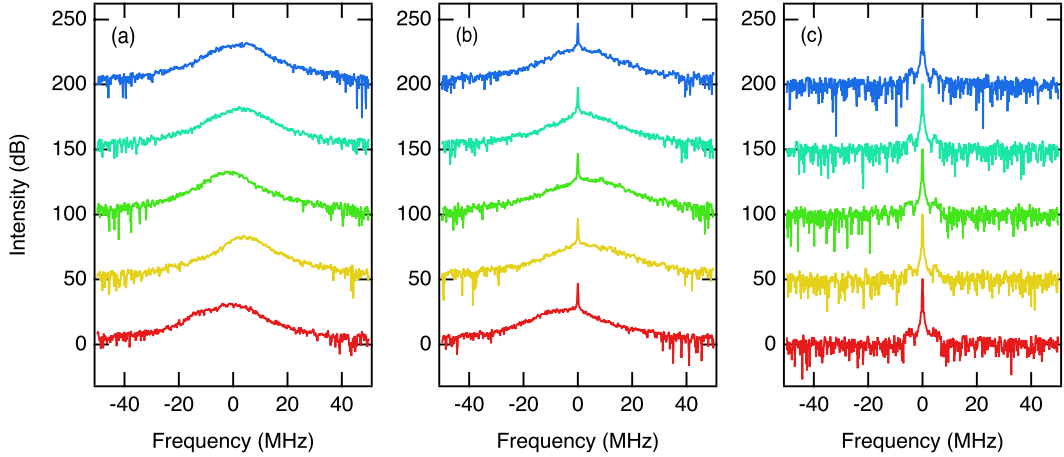

**Fig. S13** Spectra at several timings in the spectrograms of Fig. S12. Fig. S13 (a)-(c) corresponds to Fig. S12 (a)-(c).

### (3) Post-selection method

First, we explain the post-selection analysis in the case of Fig. 5 (a). Figure S14 (a) shows a bare spectrogram (10,000 traces) measured when a signal is injected close to the free-running frequency with a normalized injection amplitude of $k = 1.6 \times 10^{-2}$. The horizontal axis is the frequency offset to the injection frequency. To pick-up the spectra whose free-running frequency is the same as the injection frequency, we performed the following selection procedure: First, we calculated the spectral centroid at each moment, as shown in Fig. S14 (b). Then, we sorted the spectra by the spectral centroid, as shown in Fig. S15 (a). Figure S15 (b) shows the corresponding spectral centroid. Next, we picked-up spectra with the same spectral centroid as the injection frequency (50 traces from 10,000 traces) and averaged them to obtain the spectra shown in Fig. 5 (a). We note that this method can be extended to pick-up a spectrum with an arbitrary offset-frequency between the injection-locked component and the noise component.

In the case of Fig. 2 (a), we measured a spectrogram of the free-running RTD oscillator in the same way and picked-up the spectra with a common center frequency.

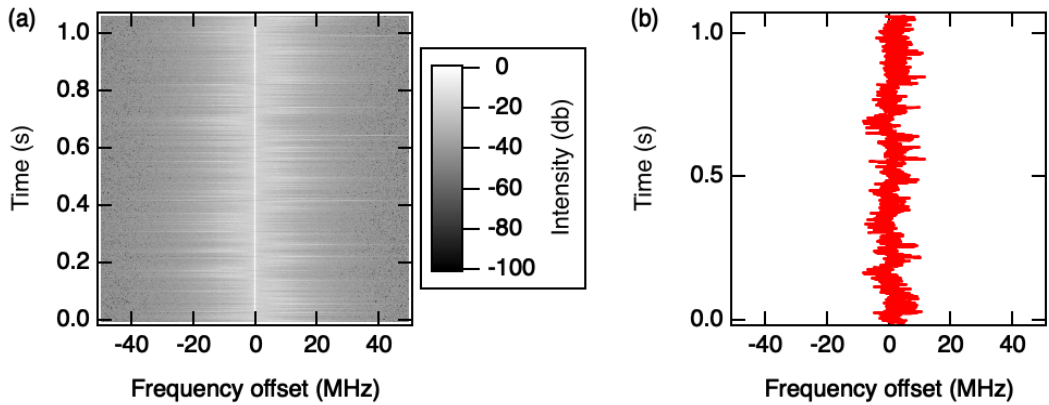

**Fig. S14** (a) Spectrogram measured when the injection frequency is close to the free-running frequency (the normalized injection amplitude $k = 1.6 \times 10^{-2}$). (b) Spectral centroid at each moment.

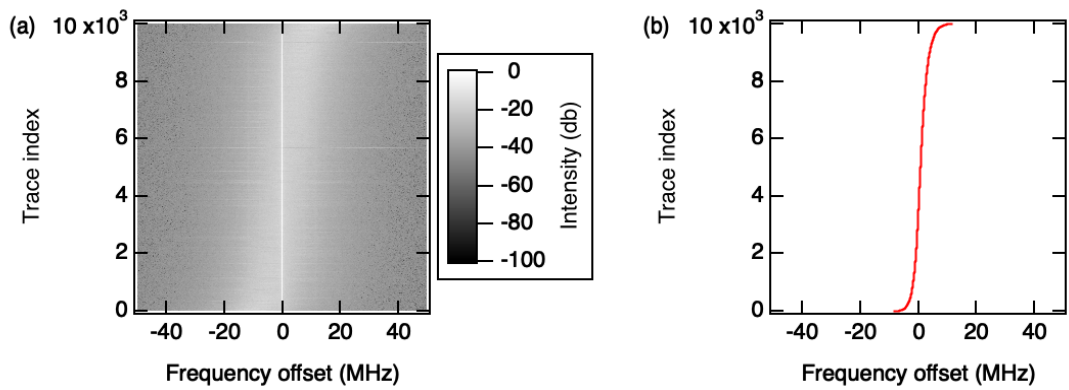

**Fig. S15** (a) Spectrogram sorted by the spectral centroid and (b) spectral centroid of each trace.

**Section 6   Measurement of injection amplitude and injection ratio**

In this section, we describe how to determine the injection amplitude $\tilde{V}_{\text{inj}}$ and injection ratio $r = \tilde{V}_{\text{inj}}/\tilde{V}_{\text{osc}}$. $\tilde{V}_{\text{inj}}$ is the injection amplitude at the antenna caused by the injection electric field. $\tilde{V}_{\text{osc}}$ is the oscillation amplitude of the RTD oscillator itself measured at the antenna. In this section, tilde on $\tilde{V}_{\text{inj}}$ and $\tilde{V}_{\text{osc}}$ are used to clarify that these symbols represent the amplitude of the ac voltage at THz frequency.

**(1) Measurement of the injection voltage ($\tilde{V}_{\text{inj}}$)**

We performed a square-law detection of the injection THz wave using the RTD THz oscillator[2,3] with a setup shown in Fig. S16. The injection THz wave was generated with a UTC-PD. The laser incident on the UTC-PD was modulated square-wave on-off shape at 9.7 MHz. The bias voltage of the RTD THz oscillator was set to 406 mV, where no oscillation took place.

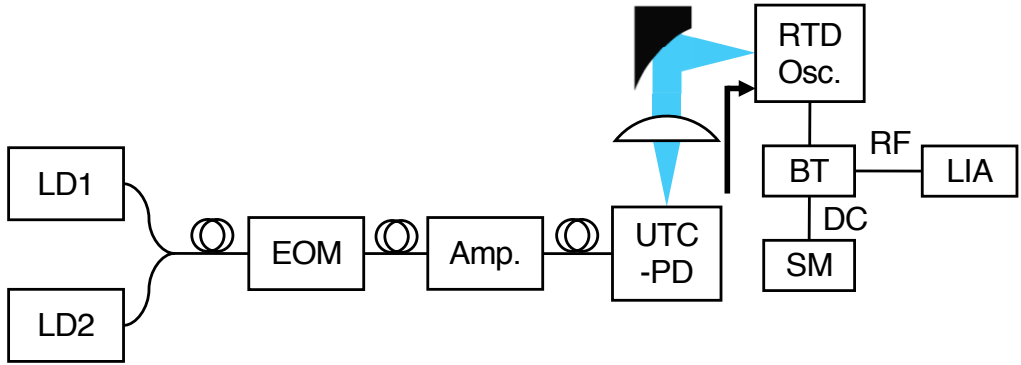

**Fig. S16** Setup for the measurement of injection voltage. The emission of two laser diodes (LD1 and LD2) are input to the electro-optic modulator (EOM), the optical amplifier (Amp.), and the UTC-PD via optical fiber. Intensity-modulated THz wave is generated with the UTC-PD and injected to the RTD THz oscillator. The RTD THz oscillator is connected to a bias tee (BT). A source meter (SM) supplies DC voltage to the RTD THz oscillator via the low-frequency (DC) port of the bias tee. The high-frequency (RF) port is connected to a lock-in amplifier (LIA) and the square-law signal is measured.

Figure S17 (a) shows the current-voltage curve $I_{\text{FP}}(V)$ in the voltage range of 300-500 mV. We can utilize its nonlinearity for THz-wave detection. When a signal of $\tilde{V}_{\text{inj}} \cos \omega t$ is injected, time-averaged current changes from that without the injection. The difference is described as

$$\Delta I_{\text{DC}}(\tilde{V}_{\text{inj}}) = \overline{I_{\text{FP}}(V_{\text{DC}} + \tilde{V}_{\text{inj}} \cos \omega t) - I_{\text{FP}}(V_{\text{DC}})}, \qquad (S6-1)$$

where $V_{\text{DC}}$ is the bias voltage, and the overline represents time average. When $\tilde{V}_{\text{inj}}$ is small, it can be calculated as

$$\Delta I_{DC}(\tilde{V}_{inj}) = \frac{1}{4}\left|I_{FP}^{(2)}(V_{DC})\right|\tilde{V}_{inj}^{2} + \frac{1}{64}\left|I_{FP}^{(4)}(V_{DC})\right|\tilde{V}_{inj}^{4} + O\left(\tilde{V}_{inj}^{6}\right) \quad (S6-2)$$

$$= \Delta I_{2}(\tilde{V}_{inj}) + \Delta I_{4}(\tilde{V}_{inj}) + O\left(\tilde{V}_{inj}^{6}\right), \quad (S6-3)$$

where $I_{FP}^{(n)}(V)$ represents the n-th order derivative of $I_{FP}(V)$.

To obtain the derivative coefficients, we fitted $I_{FP}(V)$ around the bias voltage of 406 mV. Figure S17 (b) shows $I_{FP}(V)$ as the red line in the voltage range of 397-415 mV. We fitted the current-voltage curve in the voltage range of 401-411 mV. To confirm that the higher-order term in Eq. (S6-3) does not contribute to the signal $\Delta I_{DC}$, we used the 4th-order polynomial:

$$I_{FP}(V) = I_{FP}(V_{DC}) + I_{FP}^{(1)}(V_{DC})\Delta V + \frac{1}{2!}I_{FP}^{(2)}(V_{DC})\Delta V^{2}$$
$$+ \frac{1}{3!}I_{FP}^{(3)}(V_{DC})\Delta V^{3} + \frac{1}{4!}I_{FP}^{(4)}(V_{DC})\Delta V^{4}, \quad (S6-4)$$

where $\Delta V = V - V_0$, and $V_{DC}$ is the bias voltage of 406 mV. The fitting result is shown as the dashed curve in Fig. S17 (b). Table ST2 shows the fitting parameters and standard deviations as well as $\Delta I_n(5\text{ mV})$ in Eq. (S6-3) for $n = 2$ and 4. We can see that $\Delta I_2(5\text{ mV}) \gg \Delta I_4(5\text{ mV})$. Hence, the higher order terms in Eq. (S6-3) is small in the case of $\tilde{V}_{inj} \leq 5$ mV, which holds true in our experiment as described later.

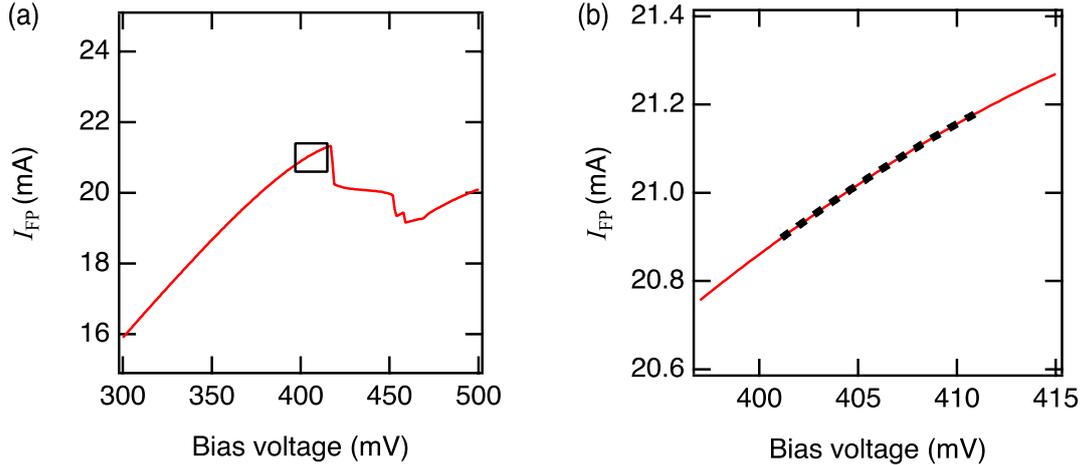

**Fig. S17** (a) Current-voltage curve $I_{FP}(V)$ in the voltage range of 300-500 mV. The boxed region corresponds to the plot range of (b). (b) Current-voltage curve $I_{FP}(V)$ in the voltage range of 397-415 mV measured with the voltage step of 100 μV (red line) and its fitting curve (black dashed line).

**Table ST2** Coefficients of the fitting on the current-voltage curve $I_{\text{FP}}(V)$ with Eq. (S6-4).

| Order $n$ | 0 | 1 | 2 | 3 | 4 |
|---|---|---|---|---|---|
| Value of $I_{\text{FP}}^{(n)}(V_{\text{DC}})$ (A/V$^n$) | $2.1 \times 10^{-2}$ | $2.9 \times 10^{-2}$ | $-8.2 \times 10^{-1}$ | $-3 \times 10$ | $-1 \times 10^4$ |
| Standard deviation of $I_{\text{FP}}^{(n)}(V_{\text{DC}})$ (A/V$^n$) | $3 \times 10^{-8}$ | $1 \times 10^{-5}$ | $1 \times 10^{-2}$ | 4 | $7 \times 10^3$ |
| $\Delta I_n(5\ \text{mV})$ (A) | - | - | $5 \times 10^{-6}$ | - | $1 \times 10^{-7}$ |

Figure S18 shows the equivalent circuit for the measurement. The RTD THz oscillator was composed of an RTD, an LCR circuit, and a MIM capacitor. The bias voltage was applied to the RTD THz oscillator through the DC port of a bias tee. We used a bias tee with a frequency range of 0.1-6000 MHz. The modulated injection field caused a modulated square-law detection signal, which is coupled to a lock-in amplifier through the RF port of the bias tee. The square-law detection current at the feed point $\Delta I_{\text{DC}}$ is represented as

$$\Delta I_{\text{DC}} = \frac{1}{4} |I_{\text{FP}}''(V_{\text{DC}})| \tilde{V}_{\text{inj}}^2, \tag{S6-5}$$

where $\tilde{V}_{\text{inj}}$ is the injection amplitude at the antenna. The voltage measured at the lock-in amplifier is described as follows:

$$V_{\text{LIA}} = R_{\text{LIA}} \Delta I_{\text{DC}} = \frac{1}{4} R_{\text{LIA}} |I_{\text{FP}}''(V_{\text{DC}})| \tilde{V}_{\text{inj}}^2. \tag{S6-6}$$

Therefore,

$$\tilde{V}_{\text{inj}} = 2\sqrt{V_{\text{LIA}}/R_{\text{LIA}}|I_{\text{FP}}''(V_{\text{DC}})|}. \tag{S6-7}$$

Related values were $R_{\text{LIA}} = 50\ \Omega$ and $I_{\text{FP}}'' = -0.82$ A/V$^2$. The maximum injection voltage ($\tilde{V}_{\text{inj,max}}$) was determined from the maximum value of $V_{\text{LIA}}$ (6.9 μV) in the series of experiments in this paper as follows:

$$\tilde{V}_{\text{inj,max}} = 0.82\ \text{mV}. \tag{S6-8}$$

We also confirmed that $V_{\text{LIA}}$ was proportional to $\tilde{V}_{\text{inj}}$ by attenuating the injection field with a pair of WGs. We note that $\tilde{V}_{\text{inj}}$ in the case of no attenuation slightly varied from experiment to experiment depending on the conditions such as optical alignment. In each experiment, we first measured $\tilde{V}_{\text{inj}}$ in the case of no attenuation, and derived other $\tilde{V}_{\text{inj}}$ values for various attenuations using the proportionality.

$\tilde{V}_{\text{inj}}$ has two significant figures because the related quantities $I_{\text{FP}}''$, $V_{\text{LIA}}$ and the transmission of the WG pair has two significant figures.

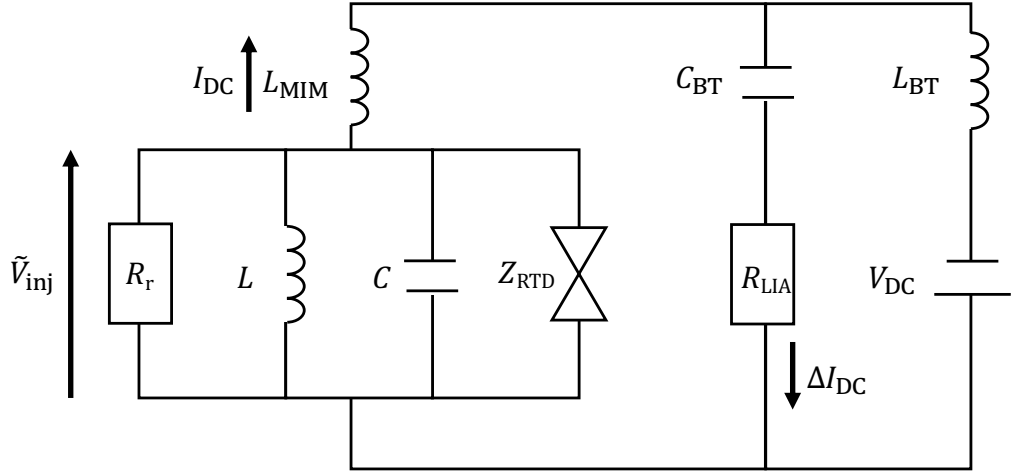

**Fig. S18** Equivalent circuit for the measurement of injection voltage. $R_r$, $L$ and $C$ are the resistance, the inductance, and the capacitance of the antenna, respectively. $Z_{RTD}$ is the impedance of the RTD. $L_{MIM}$ is the inductance of the MIM capacitor. $C_{BT}$ and $L_{BT}$ are the capacitance and the inductance of the bias tee, respectively. $R_{LIA}$ is the imput impedance of the lock-in amplifier. $V_{DC}$ is the bias voltage.

### (2) Estimation of the oscillation voltage ($\tilde{V}_{osc}$)

We estimated the oscillation voltage from the radiation power. The assumed equivalent circuit of the RTD THz oscillator is shown in Fig. S19. The RTD THz oscillator is composed of a negative resistance $-R$ and an LCR resonator, which corresponds to the antenna. The oscillation amplitude $\tilde{V}_{osc}$ can be represented as

$$\tilde{V}_{osc} = \sqrt{2 P_{out} R_r}, \qquad (S6-9)$$

where $P_{out}$ is the emission power, and $R_r$ is the radiative resistance. The emission power was typically 8 µW, as shown in Figure 1 (c). The antenna was a half-wavelength antenna. We assume its radiative resistance as 150 Ω, as in the previous study on the RTD THz oscillator of a similar structure.[2] Therefore, we can derive the oscillation voltage as

$$\tilde{V}_{osc} = 50 \text{ mV}. \qquad (S6-10)$$

Here, $\tilde{V}_{osc}$ has one significant figure in the above derivation because $P_{out}$ has only one significant figure due to noise in the measurement, and the assumption of $R_r$ has uncertainty due to the unknown effective refractive index of the substrate.

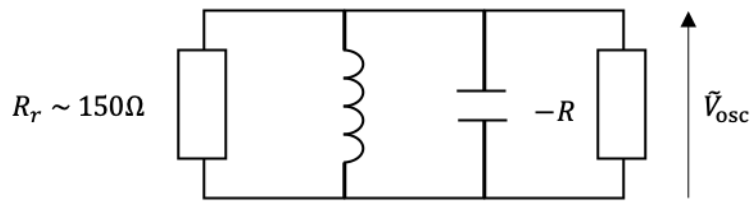

**Fig. S19**：Equivalent circuit of the RTD THz oscillator to calculate the oscillation voltage

**(3) Derivation of the injection ratio**

From the results above, we can derive the injection ratio $r = \tilde{V}_{\text{inj}}/\tilde{V}_{\text{osc}}$. Its maximum value in the series of experiments in this paper was

$$r_{\max} = \frac{\tilde{V}_{\text{inj,max}}}{\tilde{V}_{\text{osc}}} = \frac{0.82}{50} = 2 \times 10^{-2} \tag{S6 - 11}$$

An injection ratio for a normalized injection amplitude $k = \tilde{V}_{\text{inj}}/\tilde{V}_{\text{inj,max}}$ can be calculated as

$$r = \frac{\tilde{V}_{\text{inj}}}{\tilde{V}_{\text{osc}}} = \frac{k\tilde{V}_{\text{inj,max}}}{\tilde{V}_{\text{osc}}}. \tag{S6 - 12}$$

Here, $r$ has one significant figure because $\tilde{V}_{\text{osc}}$ has one significant figure, while $\tilde{V}_{\text{inj}}$ has two significant figures.

**(4) Verification of the coupling efficiency by beam shape**

So far, we derived the maximum injection voltage $\tilde{V}_{\text{inj,max}} = 0.82$ mV. The radiative resistance is 150 Ω, so it corresponds to an injection power of $P_{\text{inj,max}} = 2$ nW. The incident THz power in front of the RTD THz oscillator is $P_0 = 2$ μW. Hence, the coupling efficiency in power can be derived as

$$\frac{P_{\text{inj,max}}}{P_0} = \frac{2 \text{ nW}}{2 \text{ μW}} = 1 \times 10^{-3}. \tag{S6 - 13}$$

We verify this coupling efficiency by comparing the beam area and the effective antenna area. We estimate the beam radius focused on the antenna ($w_{03}$) with the formula for the Gausssian beam:

$$w_{03} = \left(\frac{2f}{kw_{01}}\right) \bigg/ \sqrt{1 + \left(\frac{2f}{kw_{01}}\right)^2}. \tag{S6 - 14}$$

Here, $f = 100$ mm is the focal length of the parabolic mirror, $k = 6.3$ /mm is the wavenumber, and $w_{01} = 8$ mm is the radius of the parallel beam. From this formula, we can derive that $w_{03} = 3$ mm. It is known that the effective area of a dipole antenna can be represented as

$$A_{\text{e}} = 0.13\lambda^2, \tag{S6 - 15}$$

where λ is the effective THz wavelength at the antenna. The antenna length of the RTD THz oscillator was measured as 166 μm with an optical microscope. This corresponds to the half of the effective wavelength, so the antenna is designed for $\lambda = 332$ μm and the effective area of the dipole antenna is $A_{\text{e}} = 0.014$ mm². Hence, the coupling efficiency derived from the beam shape is

$$\frac{A_{\text{e}}}{\pi w_{03}^2} = 4 \times 10^{-4}. \tag{S6 - 16}$$

This value is comparable to the value of Eq. (S6-13).

**Section 7  Circuit simulation to verify the effect of voltage-dependent capacitance**

Figure S20 shows an equivalent circuit of an RTD oscillator for the circuit simulations. We modeled the RTD as the parallel connection of the voltage-dependent resistance and the voltage-dependent capacitance. We used expressions presented by Diebold[4] for the current-voltage curve and the capacitance-voltage curve of an RTD. The circuit topology and some values of the lumped elements were derived from Ref. 2. We put a white noise source of $1.8 \times 10^{-20}$ W/Hz with a cutoff frequency of 10 THz in series with the DC voltage source to reproduce the linewidth of the emission spectrum. The differential equations for the circuit were converted to difference equations. The difference equations were solved numerically with the time transient analysis. We set the injection amplitude $V_{\text{inj}}$ as 2 mV. On the antenna resistance, it caused voltage with an amplitude of 1.9 mV in the non-oscillating condition with the bias voltage of 0.35 V. The typical oscillation amplitude on the antenna was 345 mV. Hence, the typical injection ratio here was $1.9 \text{ mV}/345 \text{ mV} = 6 \times 10^{-3}$. It is the same as the injection ratio of $r = 1 \times 10^{-2}$ in Fig. 4 (b).

We performed simulations for two cases. For the first case, capacitance of the RTD does not vary in time. It is a constant value of $C_{\text{RTD}}(V_{\text{RTD},0})$, where $C_{\text{RTD}}(V)$ is the dependence of the capacitance of the RTD on the voltage $V$, and $V_{\text{RTD},0}$ is the voltage applied on the RTD when we apply DC bias voltage with the capacitors open and the inductor shorted. For the second case, we assumed that the capacitance responds to the AC voltage on the RTD $V_{\text{RTD}}(t)$ instantaneously. Then, capacitance of the RTD is a time-varying value of $C_{\text{RTD}}(V_{\text{RTD}}(t))$. The other parameters used in the two cases were the same.

Figure S21 shows the free-running frequency, the free-running linewidth, and the locking range calculated with the constant RTD capacitance. Here, the locking range was independent of the bias voltage. Figure S22 represents the result for the case of the time-varying capacitance. Here, the locking range was bias-voltage dependent, and it was asymmetric at several bias points. Hence, these results support our statement that the voltage-dependent capacitance results in the bias-voltage dependent and asymmetric locking range.

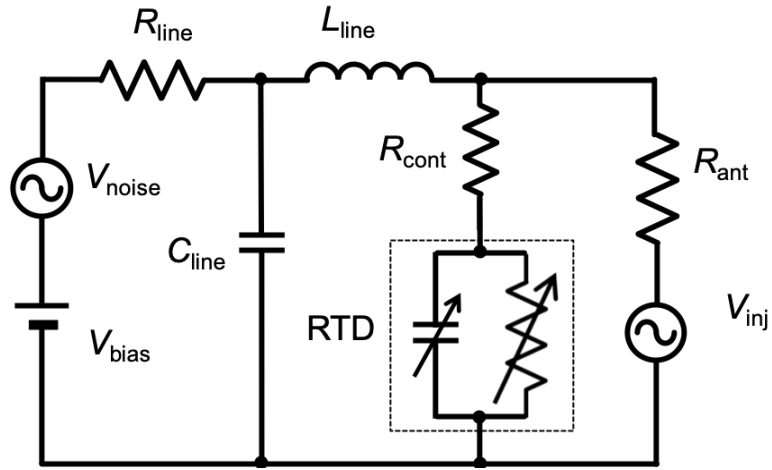

**Fig. S20** Equivalent circuit of an RTD oscillator with an injection signal source. The area of the RTD is 1.6 μm². The values of the lumped elements of $R_{ant}$ = 150 Ω, $L_{line}$ = 17 pH, $C_{line}$ = 120 fF were derived from Ref. 2. We set the values of $R_{line}$ = 5 Ω, and $R_{cont}$ = 1 Ω as a reasonable value.

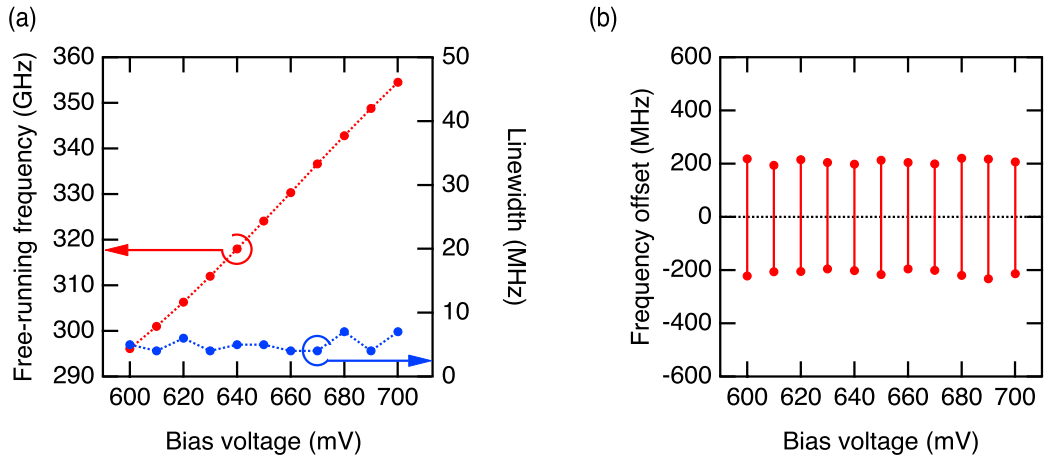

**Fig. S21** Simulation results with constant RTD capacitance. (a) Free-running frequencies and free-running linewidths at several bias voltages. (b) Locking ranges at several bias voltages for an injection amplitude of 2 mV are shown as frequency offset from the free-running frequency.

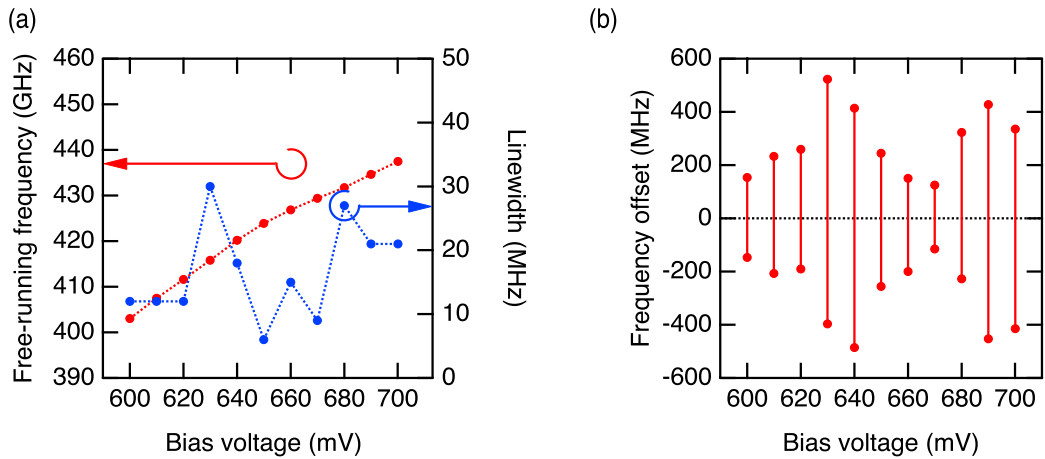

**Fig. S22** Simulation results with time-varying RTD capacitance. (a) Voltage-dependence of the free-running frequency and the free-running linewidth. (b) Voltage-dependence of the locking ranges at several bias voltages for an injection amplitude of 2 mV.

**Section 8  Peak area derivation and correction**

In this section, we show how to derive the spectral area shown in Fig. 5 (b). We fitted the narrow and broad peaks in Fig.5 (a) with the Lorentzian functions. The HWHMs of both peaks were independent of the normalized injection amplitude $k$. The HWHM of the narrow peak was about 50 kHz, which corresponds to the RBW of the spectrum analyzer. The HWHM of the broad peak was 4.4 MHz.

Here, the spectral height and the area of the broad peak were underestimated in the RTSA because of their noisiness and the frequency fluctuation. To correct this, we executed an independent total power measurement. In this measurement, we modulated the emission of the RTD THz oscillator using an optical chopper and performed a square-law detection with an FMBD. We compared the emission power in the free-running condition and the injection-locked condition. Here, the injection amplitude was the maximum value in our setup, i.e., the normalized injection amplitude $k$ was about unity. We found that the total emission power was almost the same; the power difference in the two cases was less than 2%, which was comparable to the noise level of the measurement. We also confirmed that the total emission power was constant for various injection amplitudes.

From above experiments, we determined the correction factor for the spectral area of the broad peaks as 1.7 to keep the total peak area constant. We multiplied the spectral areas of the broad peaks by the factor, and normalized all the spectral areas with the total spectral area at $k = 1$ to obtain Fig. 5 (b).

**Section 9  Effect of flicker noise on the output noise spectra**

In this section, we look into the detail of the output noise spectra (Fig. 6 (a)). We show that the spectra can be explained as a result of a flicker noise source, especially in the high frequency part, where the output noise is small compared to the signal. The flicker noise is the noise with the power spectrum of $f^{-\alpha}$ (0<α<2).[5] There are several types of theories on phase noise of the oscillator. They can be classified into the linear time-invariant (LTI) theories,[6] linear time-variant (LTV) theories,[7,8] and nonlinear time-variant (NLTV) theories.[9-11] We use the LTV theories in this section.

The oscillator's output voltage $V_\text{out}(t)$ is expressed as

$$V_\text{out}(t) = \big(A + \Delta A(t)\big) g\big(2\pi f_0 t + \Delta \phi(t)\big), \qquad (S9-1)$$

where $A$ is the amplitude, and $f_0$ is the oscillation frequency without a noise effect, and $g$ is a periodic function with a period of $2\pi$. $\Delta\phi(t)$ is the phase fluctuation, and $\Delta A(t)$ is the amplitude fluctuation due to the noise source, such as the current fluctuation. In the output noise spectrum, the amplitude noise is usually much smaller than the phase noise.[7] We ignore the amplitude noise here.

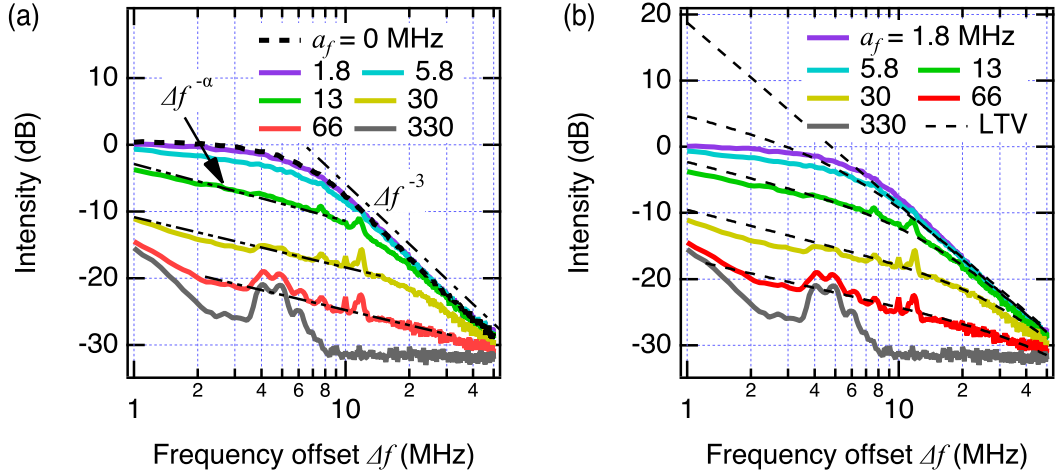

**Fig. S23** (a) Measured power spectra for various injection strength. The half locking range $a_f$ represents the injection strength. The chain line is a guide-to-the-eye indicating the slope of $f^{-3}$ in the free-running spectrum. The two-dot chain lines indicate the slope of $f^{-\alpha}$ in the injection-locked case of $a_f = 13, 30,$ and $66$ MHz. $\alpha = 0.8$ is derived from the fitting. (b) Fitting result of the measured noise spectra with Eq. (S9-4). Dashed lines are the fitting curve.

In Fig. S23 (a), we again show the noise spectra of Fig. 6 (a). In the free-running case ($a_f = 0$), the output noise spectrum has an $\Delta f^{-3}$ tail in the high frequency part and a flat region in the middle. The $\Delta f^{-3}$ slope may result from the up conversion of the input $\Delta f^{-1}$ noise to the carrier frequency by the oscillator. This effect is well-known as the Leeson effect described as[6,7]

$$L_{\phi,\text{free}}(\Delta f) = \left(\frac{f_0}{2Q\Delta f}\right)^2 L_{\phi,\text{input}}(\Delta f) \qquad (S9-2)$$

in the frequency range of $\Delta f \ll f_0/2Q$. Here, $L_{\phi,\text{free}}(\Delta f)$ is the output phase noise spectrum and $Q$ is the Q-factor of the resonator. $L_{\phi,\text{input}}(f)$ is the power spectrum of the input phase fluctuation defined as

$$L_{\phi,\text{input}}(f) = |\mathcal{F}[\Delta\phi(t)]|^2 \qquad (S9-3)$$

where $\mathcal{F}$ means the Fourier transform. We note that the variable $f$ is not the frequency offset but the frequency around DC. In Leeson's formula (S9-2), there is a problem that the output noise spectrum diverges at $\Delta f = 0$. It is because this theory can be applied only in the high frequency region where the output noise is small. Therefore, we focus only on the high frequency part where the divergence does not matter.

In the injection-locked case of $a_f = 13, 30,$ and $66$ MHz, each output noise spectrum is composed of an approximately $\Delta f^{-3}$ tail and an $\Delta f^{-\alpha}$ ($\alpha \sim 0.8$) part at the center. Here, the value of $\alpha$ is determined by the fitting with the following function:

$$L_{\phi,\text{out}}(\Delta f) = \frac{A}{\Delta f^\alpha}\frac{1}{\Delta f^2 + a_f^2} \quad (S9-4)$$

where $A$ and $\alpha$ are common parameters for these traces. The fitting curves shown in Figure 20 (b) well reproduce the experimental results. It indicates that the output power spectrum is determined by the input phase fluctuation with the power spectrum of $f^{-\alpha}$.

Eq. (S9-4) is derived from an LTV theory.[8] According to Eq. (32) in Ref. 8, output noise spectra of an injection-locked oscillator can be described as

$$L_{\phi,\text{out}}(\Delta f) = L_{\phi,\text{free}}(\Delta f)\frac{\Delta f^2}{\Delta f^2 + a_f^2} + L_{\text{ext}}(\Delta f)\frac{a_f^2}{\Delta f^2 + a_f^2}, \quad (S9-5)$$

where $L_{\phi,\text{free}}(\Delta f)$ is the free-running output noise spectrum, and $L_{\text{ext}}(\Delta f)$ is the phase noise spectrum of the injection signal. We assume the power spectrum of input phase fluctuation is

$$L_{\phi,\text{input}}(\Delta f) = \frac{A'}{\Delta f^\alpha}, \quad (S9-6)$$

where $A'$ is a constant. Then, the free-running output noise spectrum $L_{\phi,\text{free}}(\Delta f)$ expected from Leeson's formula (S9-2) is

$$L_{\phi,\text{free}}(\Delta f) = \frac{A}{\Delta f^{2+\alpha}}. \quad (S9-7)$$

Here, $A = A'(f_0/2Q\Delta f)^2$. By substituting Eq. (S9-7) to Eq. (S9-5), we obtain

$$L_{\phi,\text{out}}(\Delta f) = \frac{A}{\Delta f^\alpha}\frac{1}{\Delta f^2 + a_f^2} + L_{\text{ext}}(\Delta f)\frac{a_f^2}{\Delta f^2 + a_f^2}. \quad (S9-8)$$

By neglecting the second term in Eq. (S9-8), we can obtain Eq. (S9-4). In the strong injection case ($a_f = 330$ MHz), we cannot neglect the second term. It is because when $a_f \gg \Delta f$, the first term vanishes while the second term approaches to $L_{\text{ext}}(\Delta f)$.

Equation (S9-4) does not fit the experimental result in the case of $a_f = 1.8$ and 5.8 MHz. This is due to the problem of the output noise divergence at low frequency in the LTV theories. We can avoid the divergence with NLTV theories in the case of a free-running oscillator with a colored noise source[10] and an injection-locked oscillator with a white noise source.[11] However, NLTV theory on the injection locking under colored noise source has not been developed.

In the above discussion, there is a strange point that it looks like the power spectrum of the noise source changed from $f^{-1}$ to $f^{-0.8}$ due to the injection locking. If this was not an artifact, a possibly related mechanism is the photon-assisted tunneling[12] that changes conduction property of the RTD due to the strong THz electric field applied on it.